# Taxonomy-Driven Graph-Theoretic Framework for Manufacturing Cybersecurity Risk Modeling and Assessment

Md Habibor Rahman[a], Erfan Yazdandoost Hamedani[a], Young-Jun Son[b], Mohammed Shafae[a, 1]

[a] Department of Systems and Industrial Engineering, The University of Arizona, Tucson, AZ 85721, USA
[b] School of Industrial Engineering, Purdue University, West Lafayette, IN 47907, USA

## ABSTRACT

Identifying, analyzing, and evaluating cybersecurity risks is essential to devise effective decision-making strategies to secure critical manufacturing against potential cyberattacks. However, a manufacturing-specific quantitative approach is lacking to effectively model threat events and evaluate the unique cybersecurity risks in discrete manufacturing systems. In response, this paper introduces the first taxonomy-driven graph-theoretic model and framework to formally represent this unique cybersecurity threat landscape and identify vulnerable manufacturing assets requiring prioritized control. First, the proposed framework characterizes threat actors' techniques, tactics, and procedures using taxonomical classifications of manufacturing-specific threat attributes and integrates these attributes into cybersecurity risk modeling. This facilitates systematic generation of comprehensive and generalizable cyber-physical attack graphs for discrete manufacturing systems. Second, using the attack graph formalism, the proposed framework enables concurrent modeling and analysis of a wide variety of cybersecurity threats comprising varying attack vectors, locations, vulnerabilities, and consequences. The risk model captures the cascading attack impact of varying attack methods through different cyber and physical entities in manufacturing systems, leading to specific consequences. Then, the constructed cyber-physical attack graphs are analyzed to comprehend threat propagation through the discrete manufacturing value chain and identify potential attack paths. Third, a quantitative risk assessment approach is presented to evaluate the cybersecurity risk associated with potential attack paths. It also identifies the attack path with the maximum likelihood of success, pointing out critical manufacturing assets requiring prioritized control. Finally, the proposed risk modeling and assessment framework is demonstrated using an illustrative example.

## 1 INTRODUCTION

The convergence of digital technologies and physical manufacturing processes is transforming traditional hierarchical architecture based control of manufacturing systems into a more flexible and interconnected network architecture driven system [1], as illustrated in Fig. 1. This transformation enables data-driven operations, efficient and agile manufacturing processes, and improved system visibility, safety, and reliability [2]. In essence, the integration of digital technologies, such as the Industrial Internet of Things (IIoT), cloud and edge computing, digital twins, and artificial intelligence (AI) allows real-time monitoring and control, where individual entities cyber and physical entities interact directly with each other, enabling adaptive and decentralized decision-making and control. However, this rapidly growing interconnectivity and use of digital technologies significantly increase the potential entry points for adversaries, giving them a wider range of targets to exploit. For example, IoT devices used in smart manufacturing systems can be vulnerable due to software vulnerabilities and backdoors [3], poor identity management [4], IP misconfiguration [3], and inappropriate integration with the legacy system [5], allowing adversaries novel access to the system network by compromising them. Additionally, the interdependence of manufacturing cyber and physical assets, the large number of legacy systems run by outdated software, and openness of communications protocols due to the manufacturing assets' heterogeneity all exacerbate the risk of cyberattacks on "once isolated" manufacturing systems. Consequently, manufacturing operations and assets have become vulnerable to similar or even more significant cyber threats than Information Technology (IT) systems. Previously isolated Operational Technology (OT) devices, neither designed with security in mind nor mandated to be secure, have become part of the expanding and diverse cyberattack surface [6]. The increased accessibility to manufacturing system entities, combined with rapidly growing industrial control systems

---

vulnerabilities that surged by more than 2000% between 2018 and 2019 [7], is increasing cyberattacks' extent, likelihood, risk, and impact on today's manufacturing.

Threat actors can target specific manufacturing assets, exploit system vulnerabilities via well-designed attack vectors/methods, and cause attack consequences that result in organizational risks. Implications of these attacks can go beyond traditional cyber espionage losses to catastrophic system sabotage, causing operational downtime, equipment damage, and degraded product quality [8–11]. Recognizing the severity of such high-stakes attacks, research in manufacturing cybersecurity has been on the rise and can be broadly categorized into three focus areas: (1) general cybersecurity frameworks and manufacturing-specific attack taxonomies [6,9,12–14], (2) demonstration of potential cyberattacks on the manufacturing value chain [11,15–17], and (3) development and improvement of preventive and detective countermeasures against possible cyberattacks [10,18–20]. However, several other areas have not seen any manufacturing-specific research efforts yet. Most importantly, a manufacturing-specific quantitative approach to cyber-physical security risk modeling and assessment is currently missing.

Risk assessment refers to identifying, analyzing, evaluating, and prioritizing the security risk to operations, which is crucial for securing business drivers and enhancing production systems' resilience [21]. Periodic risk assessment enables practitioners to determine the infrastructure's vulnerability, evaluate the likelihood of threat events, and assess their impacts [22]. It also allows organizations to make informed decisions about risk mitigation by prioritizing critical security controls to mitigate significant risk attributes, enabling the devising of optimal strategies for security investment. Cybersecurity frameworks, such as the one proposed by the US National Institute of Standards and Technology (NIST) [23], and agencies such as the Cybersecurity Manufacturing Innovation Institute [24], recommend that the adoption and deployment of mitigation/defense techniques must be informed by risk assessment and quantification. Hence, current research on characterizing cybersecurity threats in manufacturing systems (e.g., using taxonomies) and developing potential defense methods (e.g., attack detection) must be complemented with work on manufacturing-specific quantitative risk modeling and assessment, which is the main contribution of this paper.

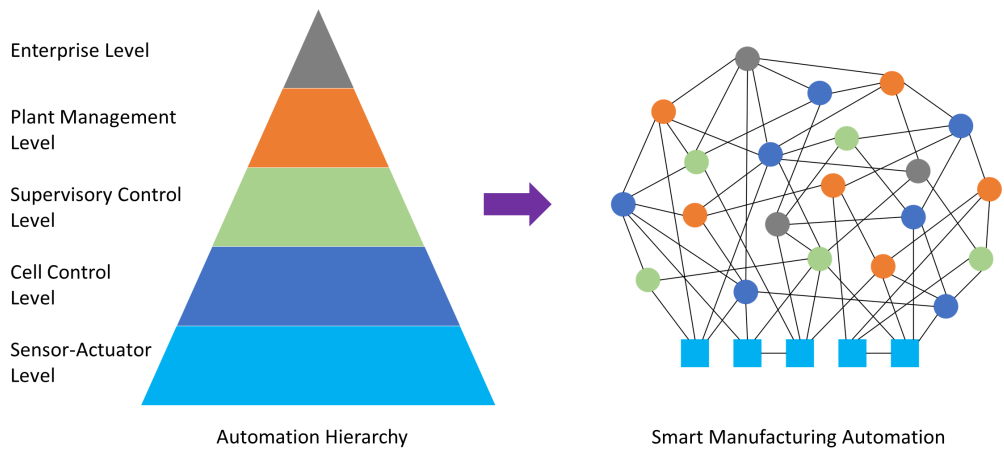

**Fig. 1.** Decomposition of centralized automation hierarchy into distributed architecture with cross-layer interactions and decentralized decision-making and control in smart manufacturing systems

To address general operational risk management needs, manufacturers commonly use qualitative risk assessment techniques such as Fault Tree Analysis (FTA) and Failure Mode and Effects Analysis (FMEA) [25,26]. However, such qualitative techniques have three major limitations in addressing the cyber-physical manufacturing security risk including: (1) being subjective and dependent on individual evaluators, (2) lack of ability to include probabilistic measures and numerical metrics for comparative analysis, and (3) inability to represent cycles such as an attacker starting at one attack location, jumping to other locations, returning to the original host, and starting in another direction. Hence, those methods are ineffective for analyzing cybersecurity risks. To address these limitations, graph-based methods can provide effective ways to model and assess the cybersecurity risk posture of cyber-physical manufacturing systems. Those methods utilize the attack graph formalism to model the interdependent system vulnerabilities and potential attack paths using network representations. By analyzing attack graphs, organizations can identify and comprehend the quantitative impact of successful attacks, prioritize risk mitigation efforts, and design adequate security controls. Attack graph-based risk assessment frameworks have been widely applied in Information Technology (IT) systems, general-purpose Industrial Control Systems (ICSs) networks, and other Cyber-Physical Systems (CPS) such as the Electrical CPS (E-CPS) [27–32]. However, there is little to no focus on developing graph-based risk management frameworks to model threat events and evaluate the unique cybersecurity risks in discrete manufacturing systems.

Discrete manufacturing systems that are being transformed into interconnected CPSs have unique characteristics compared to IT/software systems and other CPSs, such as the reliance on large numbers of legacy systems, the openness of communications protocols to connect heterogeneous manufacturing assets, diverse and expanded attack surfaces and impacts throughout large supply chains, and typically a large portfolio of products with varying product development and realization workflows and data. Therefore, risk assessment frameworks developed for IT and other CPS systems, in general, cannot accurately and effectively model and assess cybersecurity risks for smart manufacturing systems (more details are provided in Section 2.1). More specifically, current vulnerability-based graph-theoretic techniques cannot be simply extended to model and assess the cyber-physical security risks specific to discrete manufacturing systems (see Section 2.2 for details). The aforementioned unique characteristics of cyber-physical discrete manufacturing systems create a multitude of intertwined cybersecurity threats to different machines, products, and equipment, which comprise varying and distinct cyber-physical threat attributes. Adopting the attack graph formalism to model this uniquely complicated threat landscape requires the systematic generation of comprehensive and generalizable cyber-physical attack graphs. However, existing vulnerability-based risk models rely on automatic vulnerability scanning tools that only identify software and network vulnerabilities, disregarding manufacturing cyber-physical vulnerabilities in the human element, inspection systems, and production processes. Consequently, vulnerability-based attack graph generation approaches will result in incomplete and inaccurate vulnerability dependency and attack graphs, failing to offer generalizable solutions for cyber-physical manufacturing systems. Current graph-based risk assessment frameworks also focus on specific attack methods and consequences, overlooking the diverse cyber-physical attack landscape prevalent in manufacturing environments and the cascading impact of multi-stage attacks across various manufacturing assets.

To fill these research gaps, we propose a taxonomy-driven graph-theoretic framework for cybersecurity risk modeling and assessment in discrete manufacturing systems. The specific contributions of this work are summarized as follows:

1) We adopt manufacturing-specific taxonomical classifications of cyber-physical attack vectors, locations, vulnerabilities, and consequences for systematic and comprehensive characterization of the tactics, techniques, and procedures that threat actors use. Those taxonomical classifications are then used to generate comprehensive and generalizable cyber-physical attack graphs.

2) Unlike current literature mostly focusing on one or two types of attack methods and consequences, we use the attack graph formalism to model the interrelation between varying attack vectors, interdependent attack locations, and potential consequences to manufacturing assets. This enables the concurrent modeling and analysis of a wide variety of cybersecurity threats comprising varying attack attributes. By analyzing the functional dependencies among those varying threat attributes, the attack graphs can be used to explore how different attack consequences can be achieved through all applicable attack methods. In essence, the proposed graphical model facilitates the modeling of attack propagation, which refers to potential viable sequence(s) of adversarial actions comprising different attack paths through the system affecting different cyber and/or physical assets to realize the intended attack consequence.

3) We present a quantitative risk assessment model that enables estimating the likelihood of compromising manufacturing assets and probability of attack propagation, calculating the associated risks of potential attack paths, and identifying the attack path with the maximum likelihood of success. This provides a comprehensive analysis of the cascading impact of specific attack vectors targeting different assets to achieve specific consequences.

The rest of this article is organized as follows: Section 2 presents current research related to the proposed methodology and highlights their limitations. Section 3 provides details on the proposed risk modeling and assessment framework. Specifically, Section 3.1 discusses the taxonomy-driven characterization of threat attributes, the manufacturing cybersecurity risk model, its critical components and assumptions, and attack propagation. Using those building blocks, Section 3.2 presents the graph-based cyberattack modeling and attack graph representation. Section 3.3 discusses the cybersecurity risk assessment designed to assess the cascading impact of specific attack vectors targeting different assets in discrete manufacturing systems. The proposed approach is then demonstrated using an illustrative example in Section 4. Finally, Section 5 draws the paper to its conclusion.

## 2 RELATED WORKS

This section briefly discusses existing research related to the proposed methodology, including 1) graph-based risk assessment frameworks for generic IT and CPSs and 2) vulnerability-based risk modeling and attack graph generation methods. It also highlights the limitations of existing approaches in assessing cybersecurity risk in manufacturing systems.

### 2.1 Risk assessment frameworks

Graph-based risk assessment frameworks have been extensively used in modeling and assessing cybersecurity risks in IT systems, Industrial Control Systems (ICS), and other CPSs such as the power grid. Poolsappasit et al. (2011) introduced a risk assessment framework based on Bayesian Network formalism to evaluate and manage security risks in IT infrastructure [27]. Sen et al. (2016) suggested an attack graph-based risk assessment framework for heterogeneous wireless sensor networks, which also used Bayesian networks to examine and analyze attacks on the network and their impacts [28]. Ge et al. (2017) introduced a five-stage framework for assessing potential attack scenarios against the Internet of Things (IoT) using graph-based and tree-based models [29]. Huang et al. (2018) used a Bayesian Network to model the attack propagation in Industrial Cyber-Physical Systems used in chemical plants, water distribution networks, and power grids and derived the probability of compromising sensors and actuators in the system network [30]. Lyu et al. (2020) proposed a cyber-to-physical risk assessment model using a hierarchical Bayesian Network to evaluate risks in general cyber-physical systems [31]. Chae et al. (2022) proposed another risk assessment approach to assess attack paths on the instrumentation and control (I&C) systems of nuclear power plants (NPPs) [32].

However, a similar framework utilizing the attack graph formalism is missing to model threat events and evaluate the cybersecurity risk in cyber-physical manufacturing systems. Manufacturing systems have unique characteristics compared to IT/software systems and other CPS, such as the reliance on large numbers of legacy systems that are often run by outdated software, the openness of communications protocols, heterogeneous manufacturing assets, diverse attack surface including production and quality inspection processes, and manifold attack impacts [6]. Most risk assessment frameworks for ICS and general CPS are directly adapted from the cyber domain security literature, which cannot account for the unique physical characteristics of manufacturing systems and human factors. The existing graph-based frameworks will also fall short in several other aspects. For example, graph-based risk models in ICS and general CPS primarily focus on specific attack methods (e.g., Denial of Service attack) and consequences (e.g., theft of confidential information). Such model representation cannot incorporate diverse cyber-physical attack methods that can initiate threat events in manufacturing systems and the cascading impact of multi-stage cyber-physical attacks across different entities. Moreover, the graph attributes representing the threat landscape and attack propagation in manufacturing systems will fundamentally differ from ICS and other CPS. While general-purpose ICS, like the one presented in [32], often have bi-directional attack and fault propagation across the system network, cyber-physical attack paths in sequential manufacturing production processes are mostly unidirectional. The probability of successfully compromising an asset in the manufacturing-specific network topology and fault propagation from one asset to another are also different compared to ICS. Therefore, there is a pressing need to develop a risk-based framework specifically designed for the unique challenges and vulnerabilities present in manufacturing systems.

### 2.2 Vulnerability-based risk modeling and attack graphs generation

Most graph-theoretic approaches for risk modeling and assessment are vulnerability-based, where attack graphs primarily map identified system vulnerabilities automatically. Jha et al. (2002) presented an algorithm for generating attack graphs using a model-checking technique as a subroutine to analyze vulnerabilities in networked systems [33]. Ou et al. (2005) developed a software vulnerability analysis tool for networked systems relying on commonly used network vulnerability scanners for generating corresponding attack graphs for security analysis [34]. Ingols et al. (2009) evaluated the most critical threat and appropriate countermeasures in enterprise networks by creating a network model based on network vulnerability scan and firewall rules to assess the network reachability and potential attack paths for exploiting software vulnerabilities [35]. Jia et al. (2015) created a software tool to generate attack graphs and hierarchical attack representation models from vulnerability scanning reports [36]. Cai et al. (2019) proposed attack prediction and network fixing strategies for vulnerability-based attack graphs in networked systems [37].

Existing graph-theoretic methods also quantify the cybersecurity risk in terms of system vulnerabilities and emphasize vulnerability mitigation for risk reduction. Wu et al. (2016) focused on the interdependence of system vulnerabilities and used the vulnerability dependency graph to quantify the security risk in cyber-physical systems [38]. George and Thampi (2018) suggested a graph-based framework to mitigate the vulnerability exploitation risk in Industrial Internet of Things (IIoT) networks, which produced attack graphs to represent the relationships among vulnerabilities and targets in the IIoT network and measured the security threat between a given source and a target in the network [39]. Al Ghazo et al. (2019) proposed a model-checking-based automated attack graph generator and visualizer for CPS and IoT systems to analyze how threat actors can exploit interdependencies among existing vulnerabilities [40]. Ani et al. (2020) developed a Multi-Attribute Vulnerability Criticality Analysis (MAVCA) approach for estimating the impact of cyber threats and prioritizing remediation in ICSs based on three vulnerability attributes: vulnerability severities driven by environmental factors, attack probabilities relative to vulnerabilities, and functional dependencies associated with vulnerable assets [41]. Stergiopoulos et al. (2022) presented a method to analyze complex attack graphs in enterprise networks to prioritize existing Common Vulnerabilities and Exposures (CVE) vulnerabilities, analyze the impact of system states on the overall network, and suggest which combination of system states, vulnerabilities, and network configurations pose the most significant risk to the ecosystem [42].

However, vulnerability-based risk models cannot represent the holistic threat landscape in cyber-physical manufacturing systems, and the systematic generation of comprehensive attack graphs to model manufacturing cybersecurity risk is a significant challenge in utilizing the attack graph formalism. First, current approaches primarily rely on scan-based assessment tools for finding system vulnerabilities by scanning a system's network and connected devices and using these vulnerabilities for risk modeling and assessment. These automatic scanners can only identify software and network vulnerabilities in a system. In contrast, manufacturing vulnerabilities also include cyber-physical vulnerabilities in the human element, inspection system, and production process [10,11,43,44]. Therefore, existing approaches based on only cyber domain vulnerabilities and attack methods are not generalizable in manufacturing. Additionally, the type of manufacturing organization (e.g., automotive, pharmaceutical, electronics) and the specific product line can substantially impact the security risk prioritization. A manufacturer of connected IoT devices, for instance, may need to prioritize securing communication protocols and potential device vulnerabilities, whereas a manufacturer of pharmaceuticals may need to prioritize safeguarding intellectual property and ensuring the integrity of research data. Manufacturing cyber-physical threats comprised of varying threat actors and potential attack vectors cannot be modeled using the generic vulnerability-based approach that assumes both manufacturers might have the same set of cyber-physical vulnerabilities. Second, while graph-based risk assessment approaches for ICS, other CPSs, and IIoT networks primarily focus on attack prediction and risk mitigation, assuming that the attack graph will be automatically generated or is already available, generating consistent and systematic attack graphs for manufacturing systems is challenging. ICS and other CPSs, such as power grids, often have standard communication protocols, structural grid models, and power flow routines. Studies on graph-based methods for IIoT risk management also make equivalent assumptions. As a result, nodes and edges in attack graphs for the ICS, E-CPS, and IIoT networks are usually well-defined by design. In contrast, manufacturing systems have heterogeneous assets with inconsistent interconnectivity and network topology depending on the product line and service. Current attack graph generation techniques will result in incomplete and inaccurate vulnerability dependency and attack graphs due to the lack of a holistic analysis of manufacturing-specific vulnerabilities, failing to portray the convoluted risk landscape in manufacturing systems.

To address these challenges, we propose the first manufacturing-specific graph-based risk modeling and assessment framework. We use the attack graph formalism to incorporate different threat attributes and attack outcomes into the risk model for facilitating the concurrent modeling and analysis of various cybersecurity threats encompassing diverse attack vectors, locations, vulnerabilities, and consequences. Depending on the organization type and product line, manufacturers should also be able to narrow down the cybersecurity threat attributes most relevant to consider for their tailored production ecosystem. To enable such generalization, we systematically characterize manufacturing-specific cyberattack attributes using the taxonomical classification of attack attributes to represent the threat landscape and develop a graph-based risk assessment framework adapted to manufacturing systems' requirements. Our proposed framework will aid in generating comprehensive cyber-physical attack graphs for discrete manufacturing systems. This will facilitate modeling threat initiation using varying methods and the cascading attack impact through different cyber and physical entities in manufacturing systems, leading to specific consequences.

## 3 GRAPH-THEORETIC RISK MODELING AND ASSESSMENT FRAMEWORK

The proposed risk assessment approach uses attack graph formalism aiming to develop a formal model and architecture for smart manufacturing systems to (1) characterize and identify threats to operations, assets, and/or individuals, (2) model the interdependence between threat attributes and analyze the propagation of threat events, (3) formulate the likelihood of attack propagation, (4) assess cascading impacts across the value chain to evaluate the associated risk, and (5) identify the attack path with the highest risk. In line with the general guidelines for conducting risk assessments suggested by NIST [22], the proposed framework consists of three phases, as depicted in Fig. 2. The key steps are briefly outlined below.

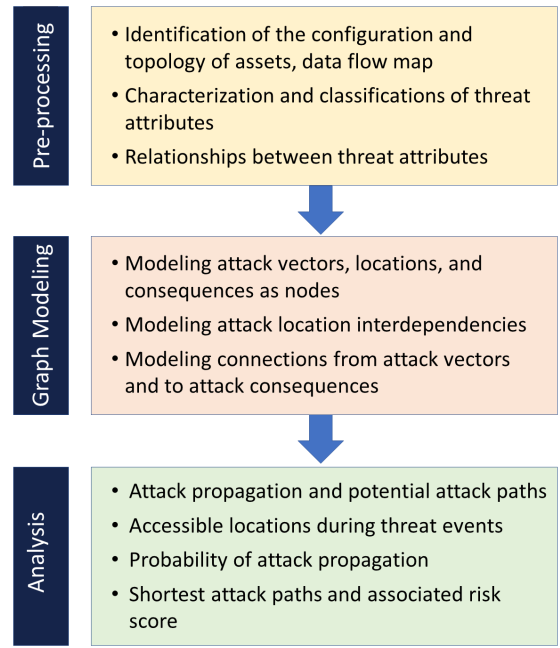

**Fig. 2.** Proposed framework for graph-theoretic risk modeling and assessment

**Pre-processing.** Current vulnerability-based risk modeling and attack graph generation approaches cannot be applied to manufacturing systems, as explained in Section 2.2. In response, this work leverages potential tactics, techniques, and procedures that threat actors can use to systematically characterize cybersecurity risk and generate attack graphs. In doing so, in the pre-processing phase, the detailed configuration of the network communication system and the topology of manufacturing assets are identified. It is necessary to distinguish all hardware, software, and operations throughout the manufacturing ecosystem, from the production floor to the corporate office and from machines to web applications. A map of the communication system and data flow within and across different assets is also necessary. Different manufacturing assets can be potential targets of threat events. In addition to the target or location of attacks, other threat attributes such as the source and the impact of those threat events should also be identified. Section 3.1 presents more details on systematically characterizing and classifying manufacturing cybersecurity threat attributes. Depending on the organizational category and product line, the relationships or interdependencies between identified threat attributes can be defined based on practitioners' opinions, domain knowledge, and/or historical data.

**Graph modeling.** The information obtained during the pre-processing phase is cognitively represented to generate attack graphs in the graph modeling phase. The identified threat attributes (attack vectors, locations, and consequences) are modeled as vertices of the graphs, whereas the edges represent their interconnections. Vertices representing attack vectors – the source of threat events – are considered source vertices in the graphical model, and the attack consequences are modeled as sink vertices. Attack locations are the candidates to which the attack vectors (source vertices) can be connected, and those vertices eventually lead to potential consequences (sink vertices). The rationale for adopting the graph-theoretic approach for risk modeling and the attack graph representation scheme are explained in Section 3.2.

**Analysis.** The generated attack graph is analyzed in the third phase. The attack propagation is mimicked using the directed edges in the attack graph. Potential attack paths from the source vertices to the sink vertices are visualized to illustrate how threat events can compromise different assets in the manufacturing ecosystem. The probability of attack propagation through different cyber and physical manufacturing assets is evaluated considering potential vulnerabilities in those assets, and the associated risk score is calculated. Finally, the shortest attack paths from different source vertices are identified based on the overall risk score, representing the most attractive attack paths for threat actors that can maximize the likelihood of success and minimize the detection probability. The shortest attack path poses the maximum risk for manufacturers and highlights the most critical manufacturing assets that need prioritized security controls and defense measures. A detailed discussion on attack propagation and risk calculation is presented in Section 3.3.

## 3.1 Taxonomy-driven modeling and characterization of manufacturing cybersecurity threats

The primary challenge to adopting the attack graph formalism for risk modeling in manufacturing systems is the systematic generation of comprehensive and generalizable cyber-physical attack graphs. This is key to enable concurrent modeling and analysis of a wide variety of cybersecurity threats comprising varying threat characteristics/attributes, which is needed to model threat propagations through the system, explore how different attack consequences can be achieved through all possible attack methods, and analyze the associated risk. To provide such a comprehensive understanding of the cybersecurity risk landscape, this work adopts the asset/impact-oriented risk assessment approach, which starts with classifying potential consequences on critical assets and identifying the threat events leading to those consequences [22]. This approach is depicted in the six-component risk model shown in Fig. 3, which combines attack consequences and the respective organizational risks with a comprehensive threat model (first four components). In general-purpose CPS risk management literature, threat events are typically defined in general terms (e.g., denial of service and system unavailability). However, given the more complicated nature of manufacturing cyber-physical attacks and their varying consequences, we use more granular information such as the Tactics, Techniques, and Procedures (TTPs) used by threat actors, following NIST's guidelines [22].

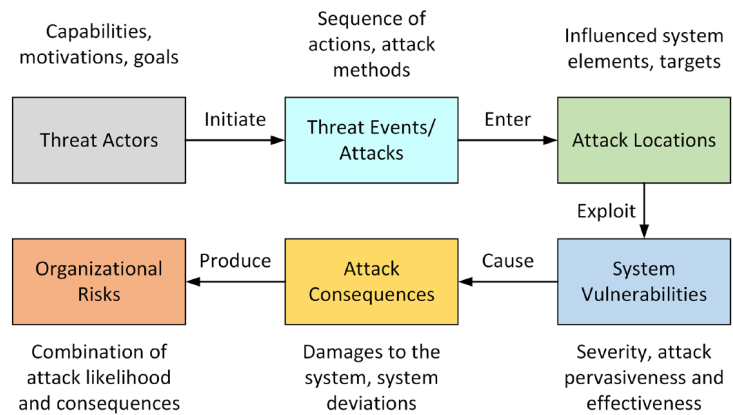

**Fig. 3.** Risk model for attacks on smart manufacturing systems

TTPs can help model and characterize manufacturing cybersecurity threats following the four-component threat model incorporated in the risk model shown in Fig. 3. Threat events encompass threat actors targeting specific locations in the manufacturing value chain, exploiting system vulnerabilities, causing system damage, and leading to organizational risk. The critical threat attributes include attack vectors, locations, vulnerabilities, and consequences. Attack vectors/methods define how threat actors can breach the system, attack locations/targets represent where threat actors can infiltrate the manufacturing value chain by exploiting certain cyber-physical vulnerabilities, and attack consequences are the effects of an attack on system assets. Attack vectors, locations, and consequences will be used to populate the attack graph nodes for risk modeling (Section 3.2) and analysis (Section 3.3). It is important to note that system vulnerabilities exist in different attack locations, which represent an opportunity for threat actors to exploit them. Knowledge of potential vulnerabilities helps determine the attack success likelihood, and this attribute is accounted for in the risk calculation.

Depending on the type of industry and product line (e.g., a defense contractor, general sheet metal parts manufacturers), organizations can narrow down the set of threat events that are most relevant to consider for their specific production environment. An organization may face a range of attack vectors, from denial-of-service attacks to advanced persistent threats. Similarly, to defend against potential attack vectors, some organizations could prioritize protecting their intellectual properties, while others could emphasize maintaining product quality and reliability. Considering the variation in potential threat events and respective attack consequences, a consistent and systematic characterization of threat attributes is essential for applying risk modeling and assessment regardless of the type of industry.

To achieve this systematic characterization, this work adopts the taxonomical classification of threat attributes derived from existing manufacturing-specific attack taxonomies to characterize manufacturing cybersecurity threats. The comprehensive, consistent, and structured classification scheme proposed in taxonomies helps to register, characterize, categorize, and communicate the cybersecurity threat attributes mentioned above. Different taxonomical classifications and respective individual elements of attack methods, locations, and consequences presented in current attack taxonomies have been compiled into a comprehensive meta-taxonomy in our previous research work [6]. The unified structure of the meta-taxonomy can guide the generation of attack graphs for manufacturers, ensuring thorough and consistent coverage of potential vertices and edges as opposed to ad-hoc experience-based populated graphs. While a detailed description of individual elements (e.g., specific attack vectors or specific system vulnerabilities) is omitted in this paper for brevity, the proposed risk model's assumptions and key components are briefly explained in the following subsections.

### 3.1.1 Threat actors

Threat actors encompass internal and external adversaries and can be defined as individuals, groups, or entities with malicious intent to compromise an organization's security. Depending on their motives, objectives, and level of sophistication, threat actors are classified into different groups, such as nation-state actors, terrorist groups, rival organizations, cybercriminals, thrill seekers, hacktivists, and insider threats [6,13,45]. Their motivations range from financial gain by stealing intellectual property and industrial espionage to political and military agendas [46]. Threat actors have varied capabilities ranging from limited domain-specific knowledge to possessing deep insights into manufacturing processes and protocols used in the control systems. Recent industry reports revealed multiple active threat groups capable of targeting industrial control systems and manufacturing entities, with new threat groups emerging every year [47]. These threat groups, such as Xenotime and Chernovite, have demonstrated their capabilities in data exfiltration and harvesting sensitive information, capturing credentials, industrial espionage, overtaking system control, compromising system safety, and disrupting multiple ICS processes, targeting manufacturers across the globe [47,48].

Additionally, insider threats pose a significant challenge in the manufacturing cybersecurity landscape compared to IT and other CPS due to the extensive human involvement throughout the product realization value chain. Malicious insiders can leverage their internal knowledge to manipulate production sequences, compromise safety measures, sabotage equipment, and/or tamper with product quality without immediate detection. Manufacturing systems often involving multiple suppliers, contractors, and partners introduce potential insider threats at various touchpoints, where insiders with access to critical information, such as proprietary designs, processes, and trade secrets, may aim to steal, replicate, or simply leak this valuable knowledge to external adversaries. Moreover, external adversaries can utilize social engineering techniques to gain insider-level access and knowledge to the system without having the insiders themselves being intentionally malicious.

In summary, manufacturing systems are susceptible to a variety of possible threat actors, ranging from opportunistic hackers to advanced persistent threat groups with substantial knowledge of industrial control systems and production processes as well as sophisticated cyber capabilities. In essence, we assume that threat actors targeting manufacturing systems will have broad capabilities and detailed knowledge of the system, and the risk modeling and assessment method proposed in this work is flexible enough to model attacks developed with varying levels of attack capabilities. Additionally, our risk model assumes that some threat actors – especially activity groups like Xenotime and Chernovite – will prioritize ensuring the success of an attack regardless of the associated cost and time required.

### 3.1.2 Attack vectors

Attack vectors or attack methods indicate the specific tools and tactics that threat actors utilize to infiltrate the manufacturing environment and perform the attack. Adversaries might use a variety of attack vectors based on their prior knowledge, access to system data, and available resources to compromise the security of a system and achieve their goals. In the context of smart manufacturing systems, attack vectors can be broadly classified into three groups depending on how the attack is executed: a) cyber domain attack vectors, b) physical attack vectors, and c) cyber-physical attack vectors. Cyber domain attack vectors, such as denial of service, malware, eavesdropping, web attack, buffer overflow, man-in-the-middle attack, replay attack, zero-day attack, and false data injection attack [49–52], can be launched over the network communication system. On the other hand, physical attack vectors, such as implanting hardware backdoors and physical tampering (e.g., de-calibrating a sensor to manipulate the input signal) [53,54], require physical access to production system equipment. A cyber-physical attack vector, such as social engineering [55], can be executed through both cyber and physical actions.

Compared to other cyber-physical systems, manufacturing systems consisting of machinery, material handling equipment (e.g., robots), in-process sensors, metrology equipment, and other physical assets have much higher heterogeneity within system components; higher product mixes leading to higher processing uncertainty; larger data volumes per production rates; and more openness to outsiders given the more complex interdependent global supply chains. Communication protocols between the heterogeneous and mostly legacy devices in manufacturing systems are typically designed to be widely open to allow greater communication flexibility, lacking appropriate security mechanisms for authentication, integrity, and measures to detect abnormal communication behavior [13]. All these factors have made manufacturing operations prone to various attack vectors. Our method considers this wide range of cross-domain attack vectors from traditional malware and denial of service attacks to manufacturing-specific attack vectors such as tampering with process parameters, which can be comprehensively identified and classified by taxonomies.

### 3.1.3 Attack locations

Attack locations or targets refer to specific elements, processes, or components within the manufacturing ecosystem that are vulnerable to cyberattacks, which threat actors seek to exploit or gain unauthorized access. Attack locations include operating systems and software, firmware, network communication systems, cloud storage, sensors, machines, products, production processes, inspection systems, human operators, and different supply chain entities [6,8,11,16,56,57]. Previously isolated OT devices designed without cybersecurity in mind are now cyber-accessible and have also become part of the expanding and diverse attack surface [6]. Different threat groups targeting manufacturers also use ICS-specific protocols for reconnaissance, accession, manipulation, and disabling of programmable logic controllers across multiple sensors and IoT devices [47]. Adversaries may gather information about the target manufacturing system and potential attack locations in the manufacturing value chain through research and reconnaissance, open-source intelligence gathering, network scanning, exploiting social engineering tactics, and leveraging insider knowledge and shared information across supply chain partners [10,11,13,47]. Hence, the proposed use of attack taxonomies, including taxonomies of manufacturing-specific attach locations, helps to account for this diverse cyber-physical attack surface in manufacturing systems.

### 3.1.4 System vulnerabilities

Vulnerabilities are inherent weaknesses or gaps within cyber-physical manufacturing systems' design, architecture, processes, or components that threat actors could exploit to compromise the system's security, integrity, functionality, and/or safety. While IT and other CPS are primarily concerned with software and network vulnerabilities, manufacturing vulnerabilities also involve cyber-physical vulnerabilities in production processes, inspection systems, and the human element [10,11,43,44]. For example, Elhabashy et al. (2020) presented several vulnerabilities in quality inspection systems, such as the improper implementation of Quality Control (QC) tools, violation of statistical assumptions of QC tools such as control charts, inadequate data collection for inspection, and inspection of a subset of product features [43]. Those QC systems vulnerabilities present new cyber-physical attack surfaces to manufacturing. In essence, those vulnerabilities can be used to design attacks on product quality integrity without being detected using typical QC tools. For example, threat actors can tamper with the Geometric Dimensioning and Tolerancing (GD&T) information used for post-production inspection, allowing the production of non-conforming products or discarding compliant ones. Moreover, other cyber-physical vulnerabilities exist in production processes and the human element [58–60]. For example, in production systems, widely used programmable logic controllers, such as the Siemens SPPA-T3000 distributed control system, often lack integrated security features and can be vulnerable due to improper authentication and cleartext transmission of sensitive information [61]. Additionally, employees working in a manufacturing system can be exploited by threat actors after they fall victim to phishing emails and/or fail to assess the risk of cyberattacks due to a lack of awareness, training, and best practices [59]. Knowledge of these cyber-physical vulnerabilities, in addition to typical IT, software, and network vulnerabilities, will help determine the attack success likelihood, and this attribute is accounted for in the risk calculation in our method.

### 3.1.5 Attack consequences

In manufacturing systems, cyberattacks can threaten data confidentiality, integrity, and availability (also known as the CIA triad in IT security), leading to extortion and intellectual property theft. The little-to-no tolerance for downtime makes manufacturers a significant target for extortion. Recent industry reports revealed that extortion and data theft were primary attack consequences to manufacturing organizations in 32% and 19% of the attack incidents, respectively [62]. Additionally, in the manufacturing physical domain, the aims of cyberattacks can go beyond traditional cyber espionage to system sabotage, causing machine breakdown, operational downtime, and compromised product quality and reliability [8,11,43,58]. For example, a cyberattack in 2022 that targeted one of Toyota's plastic part and electronic component suppliers resulted in the suspension of operations in twenty-eight production lines across Toyota's fourteen factories in Japan and the interruption of one-third of the company's global production for more than a day [63]. Several national institutes and organizations predict that manufacturing cyberattacks will increasingly focus on advanced persistent threats and sabotage rather than traditional cyber espionage to disrupt critical manufacturing ecosystems [24]. Our risk model considers all these cyber and physical domain attack consequences comprehensively categorized by different attack taxonomies.

## 3.2 Graph-based cyberattack modeling and representation

In this work, we propose a graph-theoretic cybersecurity risk modeling approach to create a formal model and architecture for representing the cyber-physical security threat landscape in manufacturing systems. Current literature mostly focuses on one or two types of attack methods and consequences and then generates attack graphs to model the linkage between relevant vulnerabilities at networked devices/systems, which may not portray the holistic threat landscape. In essence, those vulnerabilities at different system assets are modeled as graph nodes, and their dependencies are denoted as graph edges. Such representation can offer insights into threat propagation and help identify key vulnerabilities (also known as network hot spots) for specific attack methods and targets. However, it cannot account for the threat initiation using varying methods or model the cascading attack impact through cyber and physical entities of manufacturing systems, leading to specific consequences. In contrast, our proposed framework uses the attack graph formalism to model the interrelation between attack vectors, interdependent attack locations, and potential consequences to manufacturing assets, as depicted in Fig. 4. By incorporating different attack attributes in the risk model, this framework enables the concurrent modeling and analysis of a wide variety of cybersecurity threats comprising varying attack vectors, locations, vulnerabilities, and consequences. By analyzing the functional dependencies among those varying threat attributes, the graphical attack model shown in Fig. 4 can be used to explore how different attack consequences can be achieved through all applicable attack methods. Moreover, the proposed graphical model facilitates the modeling of attack propagation across diverse entities in the manufacturing value chain and depicts potential attack paths to identify the associated risks.

Attack propagation can be defined as potential viable sequence(s) of adversarial actions comprising attack path(s) through the system affecting several cyber and/or physical assets to realize the intended attack consequence. In both real-world incidents and academic literature, manufacturing cyberattacks typically involve threat actors often targeting multiple assets (attack locations) and initiating a series of attacks (sequentially and sometimes partially in tandem) that build on each other, gradually increasing their level of access and control. For example, consider that an adversary is aiming to compromise the outgoing quality and reliability of a machined part while preventing its detection. In doing so, first, they can exploit insiders (e.g., designers and machine operators) using social engineering attacks and capture security credentials through cyber domain attack vectors such as phishing and man-in-the-middle attacks. Then, the stolen security credential can be used to access the product design file or the G-code in the CNC machine controller and tamper with them to slightly alter specific dimensions in the part to degrade the part's tensile strength or fatigue life [10,11]. Additionally, the adversary can use the stolen credentials to acquire the necessary knowledge about the quality inspection strategy of the manufacturer and utilize that information to evade detection by intentionally changing the dimensions of features that are not inspected [11]. In this example, tampering with the design file or process parameters alone will not be enough to attack the product integrity because the physical manifestation of the attack may be detected through post-production inspection. Similarly, gaining knowledge of the manufacturer's inspection strategy alone will also be futile to compromise product quality and reliability without tampering with the product design and/or process. Therefore, effective manufacturing attacks will require propagating through an attack path of several cyber and/or physical assets instead of attacking a single node to realize particular objectives, and there can be multiple attack paths to realize their adversarial intent, each having different associated likelihoods. By understanding and quantifying the process of attack propagation, presented in Section 3.3, appropriate attack prevention, detection, and mitigation measures can be implemented to protect manufacturing assets and/or minimize the damage caused by cyberattacks.

In our suggested graphical model, there are three distinct groups of vertices/nodes representing attack vectors, locations, and consequences. Note that nodes and vertices are interchangeably used in graph models. The directed edges in the graph can portray the information (cyber) and material (physical) flow (i.e., cyber-physical interconnectivity among different vertices). The taxonomical classifications of attack attributes can be leveraged to populate the attack graph. In reality, a single attack vector can simultaneously target several manufacturing assets, and one asset can be affected by multiple attack vectors. For example, a false data injection attack can infiltrate the network communication system as well as the cloud storage. The network communication system can also be targeted by denial-of-service attacks. Additionally, due to the interdependence of assets, one compromised attack location can enable compromising multiple other locations in the digital manufacturing system. For example, post-production inspection systems often compare the observed quality characteristics of products (e.g., dimensions and position of part features) against standard specifications, which could be stored in the cloud. The inspection system can be compromised if threat actors gain access to the cloud storage and tamper with the standard specifications. The proposed risk modeling approach can represent all these relationships, overcoming the limitations of traditional hierarchical unidirectional and independent event tree-based risk assessment techniques. Depending on specific organizational preferences, the graphical model can be further extended to make it more granular and specific to individual assets.

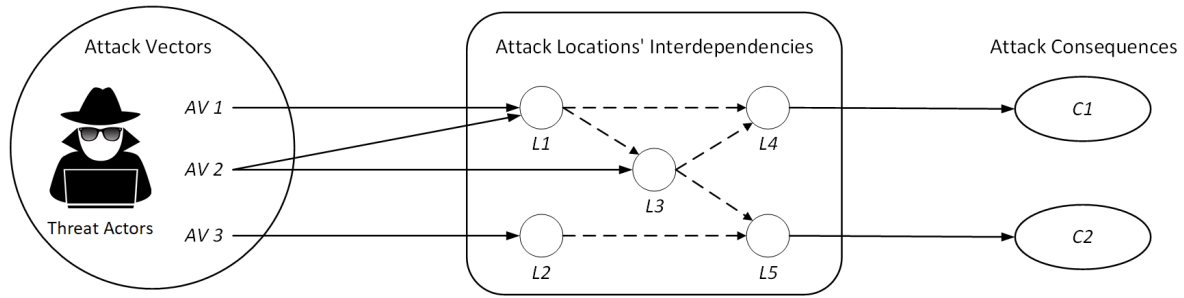

**Fig. 4.** Graph-theoretic model of cyberattacks on manufacturing systems

To illustrate, the graphical models can have a vertex representing the production process, which is categorized as a physical domain attack location as proposed in several existing taxonomies [9,11,64]. Threat actors, for example, can target this attack location and maliciously tamper process parameters (e.g., increase cutting speed in a milling machine) or digitally manipulate the machine code (e.g., offset the machine coordinate system in a CNC machine). For a more granular analysis, production processes can be classified into machining (subtractive manufacturing), additive manufacturing, joining, and assembly processes. Machining processes can be further decomposed into turning, drilling, milling, boring, laser cutting, and more. Similarly, additive manufacturing, joining, and assembly processes can be classified into sub-levels. With the necessary decomposition required for an organization, all these individual processes can be considered separate vertices in the graphical model. Such detailed decomposition can be valuable considering the varying nature of attack targets and consequences based on the process type. This will help, for example, model the propagation of a multi-stage attack to manipulate the product quality through different stages of manufacturing. Similarly, all other attack attributes can be expanded to populate the vertices in the attack graph, and respective edges can be added to represent their interconnections.

**Attack graph representation.** The attack graph presented in Fig. 4 is a directed graph consisting of a set of vertices connected by edges (also known as arcs). In this work, the attack graph will be expressed as a triplet:

$$AG = (V, E, W)$$

i. $V \subset \{1, ..., N\}$ is the set of $N$ vertices satisfying $V = \{V_{AV}\} \cup \{V_L\} \cup \{V_C\}$, where $V_{AV}$ are the vertices representing attack vectors used by threat actors to launch an attack. Potential attack locations are represented using $V_L$, and $V_C$ are the vertices depicting attack consequences. To facilitate the notation, nodes are denoted by lower-case $v$ and a subscript denoting its label. $v_{av_i} \in V_{AV}$ denotes the $i^{th}$ attack vector, $i \subset \{1, ..., N_{AV}\}$; $v_{l_j} \in V_L$ denotes the $j^{th}$ attack location, $j \subset \{1, ..., N_L\}$; $v_{c_k} \in V_C$ denotes the $k^{th}$ attack consequence, $k \subset \{1, ..., N_C\}$; where $N = N_{AV} + N_L + N_C$. To illustrate, $V = \{v_{av_1}, v_{l_1}, v_{l_2}, v_{l_3}, v_{l_4}, v_{c_1}\}$ in the directed graph presented in Fig. 5, where $v_{av_1} \in V_{AV}$ is the source of threat events, $v_{l_1}, v_{l_2}, v_{l_3}, v_{l_4} \in V_L$ are potential attack locations or manufacturing assets that could be compromised in various threat events, and $v_{c_1} \in V_C$ is a potential consequence of those threat events.

ii. $E \subset V \times V$ is the set of directed edges representing connections between different vertices. $v_{av_i} v_{l_j} \in E$ indicates that the asset $v_{l_j}$ can be accessed using $v_{av_i}$ attack vector. For the graph depicted in Fig. 5, $E = \{v_{av_1} v_{l_1}, v_{av_1} v_{l_2}, v_{l_1} v_{l_4}, v_{l_1} v_{c_1}, v_{l_2} v_{l_3}, v_{l_2} v_{c_1}, v_{l_3} v_{c_1}, v_{l_4} v_{c_1}\}$.

iii. $W \subset R^N \times R^N$ represents the weight of different edges. The weight of the edge is analogous to the distance between the two connected vertices, representing how long it will take or how difficult it is to reach from one vertex to the other in the given direction. To illustrate, in Fig. 5, $W_{AV_1,L_2} = 2$ and assume that it takes 2 hours of effort to reach $v_{l_2}$ from $v_{av_1}$ which is longer than $W_{AV_1,L_1}$ (1). In this work, $W_{AV_1,L_1}$ defines how easily threat actors can navigate from vertex $v_{av_1}$ to another vertex $v_{l_1}$, which is characterized as the inverse of the exploitation probability. The rationale is that the higher the probability of compromising an attack location by a specific attack vector is, the lower the weight of the edge connecting the attack vector to that attack location will be. As a result, it will be easier to navigate that connection for threat actors during threat events. For example, accessing $v_{l_1}$ from $v_{av_1}$ requires less resources than accessing $v_{l_2}$, i.e., compromising $v_{l_1}$ is easier (associated with higher likelihood) than compromising $v_{l_2}$ using the attack vector $v_{av_1}$. More details on calculating the probability of compromising an attack location are discussed in Sections 3.3.2 and 3.3.3.

There are two common ways to numerically represent weighted graphs for analysis. A weighted graph can be defined as (1) an adjacency list or (2) an adjacency matrix. In adjacency list representation, vertices are labeled from 1 to $|V(G)|$, and there is a list corresponding to each vertex consisting of two values. The first value denotes the destination vertex, and the second represents the weight between these two vertices. On the other hand, the graph can be represented as $AG = |V(G)| \times |V(G)|$ matrix in the adjacency matrix representation. $AG(a, b)$ element in the matrix represents the weight of the directed edge between two nodes $a$ to $b$, where a zero value represents that no edge is present between the two vertices. Note that the adjacency list representation is computationally efficient for sparse graphs where a large number of disconnected vertex pairs are present in the graph. Adding new vertices and edges to update the graph is also easier with the adjacency list. Considering the presence of heterogeneous systems and numerous potential vertices in smart manufacturing systems, adjacency list representation is used for graph representation in this work.

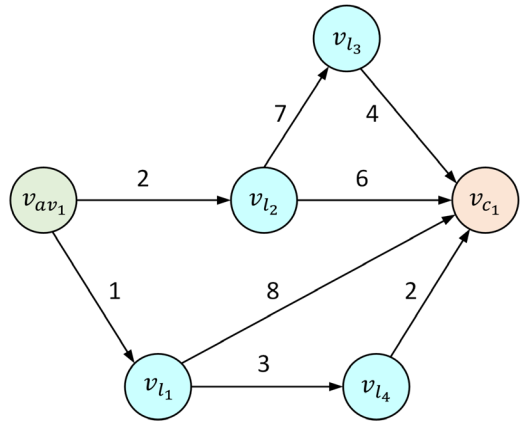

**Fig. 5.** A sample directed graph

## 3.3 Cybersecurity risk assessment

Manufacturing cybersecurity risk is calculated as a function of attack likelihood and consequences. Following the risk model presented in Fig. 3, a successful attack requires a sequence of activities from attack launch to inflicting damages to the system. In generic ICS and other CPS, attack propagation is often bidirectional; fault can propagate in either direction between two graph nodes. For example, in the power grid, power flow in transmission lines can be from generators to the load centers and vice versa. In contrast, information and material flow in multi-stage production processes are often directed, affecting fault and attack propagation. Depending on the organization type and product line, manufacturing organizations have different attack surfaces and potential attack vectors relevant to their operations. The proposed approach generates attack paths originating at specific attack vectors and then analyzes the attack propagation. Specific attack vectors can also have different cascading attack impacts depending on the location of the targeted asset in the manufacturing value chain, which is considered in the proposed risk assessment approach. The following subsections discuss the visualization of attack propagation, identification of potential attack paths, formulation of the likelihood of compromising manufacturing assets and probability of attack propagation through the digital manufacturing system, and calculation of the associated risk.

### 3.3.1 Attack propagation and potential attack paths

A set of vertices $V$ and their interconnections $E$ are required to generate the initial attack graph for understanding attack propagation through the manufacturing value chain. As mentioned before, taxonomical classifications for threat attributes can be utilized to populate the vertices, whereas expert opinion, domain knowledge, and historical data can help construct the graph edges. After creating the attack graph model, attack propagation, following different potential attack paths, will be visualized using the Depth-First Search (DFS) algorithm [65]. The DFS algorithm is often initiated at random vertices to find interconnected vertices and explore the overall connectivity in an undirected graph. However, initiating the algorithm at random vertices in the attack graph is impractical. The attack graph is a directed graph where the consequences are defined as sink vertices ($v_c \in V_C$). The exploration of attack paths starting from a sink vertex, i.e., the endpoint of a potential attack path, is not realistic. On the other hand, initiating the algorithm at vertices representing manufacturing assets ($v_l \in V_L$) could be valuable to visualize how attacks can further propagate down the manufacturing value chain but will not represent the entire attack path starting from an attack vector ($v_{av} \in V_{AV}$).

In response, this work specifies the individual starting vertex and restricts the recurrent exploration for analyzing the attack path originating from specific attack vectors. This way, the algorithm only searches for attack paths starting at the given vertex and finds how the attack can lead to potential consequences. The pseudo-code of the attack graph generation algorithm and a brief explanation are provided below.

| Depth First Search (DFS) algorithm for attack graph generation | |
|---|---|
| **Input:** | Adjacency matrix: $G$, Starting vertex $v_{av}$ |
| **Output:** | Attack propagation paths and exploited vertices |
| 1 | ***Initialize DFS*** $(G, v_{av})$, mark $v_{av}$ as visited |
| 2 | Stack $S := \{\}$ |
| 3 | *for* each vertex $u$, set $visited[u] := false$ |
| 4 | push $S, v_{av}$ |
| 5 | *while* S is not empty, *do* |
| 6 | $u := $pop S |
| 7 | *if not visited* $[u]$ |
| 8 | $visited[u] :=$true |
| 9 | *if* $u$ is in $V_{AV}$ |
| 10 | *for* each unvisited neighbor $z$ of $u$ in $V_L$ |
| 11 | push $S, z$ |
| 12 | *if* $S$ is empty |
| 13 | *for* each unvisited neighbor $z$ of $u$ in $V_C$ |
| 14 | push $S, z$ |
| 15 | *else if* u is in $V_L$ |
| 16 | *for* each unvisited neighbor $z$ of $u$ |
| 17 | push $S, z$ |

Line 1. Initializes the algorithm; the starting vertex $v_{av} \in V_{AV}$ is marked as the only visited vertex.

Lines 2 through 4. Initially, all other vertices $u \in V$ are unvisited. A stack is created to push onto the connected unvisited neighbors $u$ of $v_{av}$.

Lines 5 through 17. The iterative process continues exploring connected vertices from neighbor to neighbor – starting from attack vectors, moving into attack locations, and searching for attack consequences – before backtracking to the original vertex. Each explored vertex is marked visited, and these unvisited connected vertices are gradually added to the *visited* Boolean array. The process ends with a tree connecting all reachable vertices from the starting vertex.

The DFS algorithm begins at the starting vertex $v_{av}$, and inspects the neighbor of $v_{av}$ that has the smallest vertex index, i.e., the nearest connected neighbor to the starting vertex based on the smallest weight. Then, for the newly discovered neighbor, it inspects the next undiscovered neighbor with the lowest index. This continues until the search encounters a vertex whose neighbors have all been visited. The DFS follows the connecting edges and does not visit any vertex twice. In doing this, the algorithm keeps track of vertices that have already been visited and the path leading to the current position so that it can backtrack after all connected vertices are visited. A Boolean array keeps track of the visited vertices, and a stack is used to push vertices onto unvisited vertices. The DFS algorithm has a runtime of $O(1)$ per vertex and $O(1)$ per edge. Therefore, the time complexity of this algorithm is $O(|V| + |E|)$ when implemented using the adjacency list graph representation. The DFS algorithm is efficient and robust, which also works for disconnected digraphs, i.e., when the graph has multiple components.

### 3.3.2 Attack success likelihood

Threat actors can access the connected manufacturing systems through one or more potential entry points. The probability of compromising a specific attack location $v_{l_j} \in V_L$ using attack vector $v_{av_i} \in V_{AV}$ is calculated as the following:

$$P_{AV_i, L_j} = AV \times AC \times PR \times UI \times \frac{1}{RL}$$

Here, $P_{AV_i, L_j}$ denotes the probability of compromising the $j^{\text{th}}$ attack location using $i^{\text{th}}$ attack vector, which is a function of several metrics that collectively determine how vulnerable an attack location is to be exploited by specific attack vectors considering both cyber and physical vulnerabilities. $AV, AC, PR, UI$, and $RL$ represent the attack vector, attack complexity, privilege required, user interaction, and remediation level metrics, respectively. The $AV$ metric will have larger values for attacks that threat actors can remotely launch to compromise manufacturing assets. For example, the number of threat actors capable of launching attacks on the network communication system is higher than potential threat actors who can get physical access to a device and tamper with it. Hence, a network-level attack vector (such as a Denial-of-Service attack by sending malicious TCP packets across a network) is assigned a higher score than a physical attack vector (such as an attack using USB Direct Memory Access). The $AC$ metric represents the required control for threat actors to compromise the targeted asset; the highest score is assigned to the least complex attacks. The level of privilege that threat actors must acquire is reflected in the $PR$ metric; higher scores are assigned for significant privilege requirements such as admin control to change device settings. The $UI$ metric captures the requirement of involvement of other personnel (such as inadvertent participation of one or more employees is needed in a social engineering attack) apart from the threat actor. An asset is easier to compromise if no user interaction is required and, therefore, will have a higher $UI$ score. Finally, the $RL$ metric considers the remediation effort. The presence of cyber and physical protective and detective defense measures (such as firewall and/or monitoring process dynamics in real-time), i.e., a higher $RL$ score, reduces the probability of compromising manufacturing assets. Organizations can define the specific scoring system for those metrics as well as adopt an external scoring system such as NIST's Common Vulnerability Scoring System (CVSS) [66]. Note that $1/P_{AV_i, L_j}$ denotes the weight of the edge from any vertex $v_{av_i} \in V_{AV}$ to vertex $v_{l_j} \in V_L$, i.e., $W_{av_i, l_j}$. The lower the value of $W_{av_i, l_j}$, the easier it would be for threat actors to reach (compromise) vertex $v_{l_j}$ from vertex $v_{av_i}$.

### 3.3.3 Attack propagation likelihood

A successful attack consists of a sequence of activities from targeting an attack location, attack propagation through diverse manufacturing entities, and eventually resulting in undesirable consequences [13]. In practice, multiple attack paths can lead to the same adversarial impact. For example, a coordinated attack targeting product quality degradation may require the following steps for threat actors: a) gain access to design files, b) alter the product dimension, and c) spoof or bypass the inspection system. The same malicious intent can be realized by an insider tampering with the physical process control and sensor data. These two attack strategies require different action propagations, involve different entities in the manufacturing value chain, and will have different success probabilities. As an illustration, there are four potential attack paths leading to the attack consequence for the graph shown in Fig. 5: $v_{av_1} \rightarrow v_{l_1} \rightarrow v_{c_1}$, $v_{av_1} \rightarrow v_{l_1} \rightarrow v_{l_4} \rightarrow v_{c_1}$, $v_{av_1} \rightarrow v_{l_2} \rightarrow v_{c_1}$, and $v_{av_1} \rightarrow v_{l_2} \rightarrow v_{l_3} \rightarrow v_{c_1}$. The DFS algorithm can only discover these four attack paths without considering the likelihood of attack propagation through these potential attack paths. Once threat actors compromise an attack location, the probability of attack propagation from one vertex to another along the attack path can be modeled as the following:

$$T_{m,m+1} = Pr(x_{m+1}|x_m) \qquad m \in V_L, V_C$$

Here, $x_m$ represents that vertex $m$ has been compromised, and $T_{m,m+1}$ represents the fault propagation probability from $m^{th}$ vertex to $(m+1)^{th}$ vertex in the attack path. While this work considers $T_{m,m+1}$ as deterministic, the formulation also allows updating the transition probability. In future works, the likelihood can be dynamically updated based on gained knowledge on adversarial strategies for dynamic risk assessment. For example, the probability can be updated based on observed sequences in previous attacks, process mining and alert correlation, and post-attack forensics. Note that $1/T_{m,m+1}$ defines the weight $W_{m,m+1}$ of the edge $v_m v_{m+1}$, where $m \in V_L, V_C$. The transition probability follows the Markovian process, where the exploitation of an asset relies on the exploitation of the previous asset within the attack path. Newly discovered vulnerabilities (also known as zero-day vulnerabilities [67]) can increase the likelihood. On the other hand, the probability decreases when preventive or detective defense measures are present for monitoring an asset. For example, the manufacturer may deploy a side channel monitoring technique to observe physical process dynamics variables (such as power consumption, acoustics, and vibration) alongside the traditional post-production inspection. The side channel monitoring method can still detect anomalies in the product or process that could evade detection from the regular inspection system, reducing the probability of compromising the system [10].

### 3.3.4 Most attractive attack path identification

By combining potential attack paths and the associated likelihood of navigating each step, threat actors may want to execute the attack with minimum interference with the system to minimize the detection probability. In doing so, adversaries with the knowledge of the target system will prefer compromising the system using the attack path that maximizes the likelihood of success. As mentioned above, the weight of edges in the attack graph is inversely proportional to the likelihood of compromising the corresponding asset (viz. low edge weight denotes high likelihood). Hence, the fundamental strategy for threat actors is to follow the attack path with the lower cumulative weight in the attack graph. This work defines the attack path from a designated attack vector to a target consequence with the lowest cumulative weight as the shortest path, the most attractive path for threat actors. For example, the shortest attack path between $v_{av_1}$ and $v_{c_1}$ for the graph presented in Fig. 5 is $v_{av_1} \rightarrow v_{l_1} \rightarrow v_{l_4} \rightarrow v_{c_1}$ (shown in **Fig. 6**). Note that the shortest path does not necessarily mean the path with the smallest number of steps (hop length). While both $v_{av_1} \rightarrow v_{l_1} \rightarrow v_{c_1}$ and $v_{av_1} \rightarrow v_{l_2} \rightarrow v_{c_1}$ require only two steps – with low probabilities of exploiting $v_{c_1}$ from $v_{l_1}$ and $v_{l_2}$ (0.125 and 0.167 respectively) – the cumulative weights for these two paths are 9 and 8, respectively. Please note that the probabilities are inverse to their corresponding edge weights. In contrast, $v_{av_1} \rightarrow v_{l_1} \rightarrow v_{l_4} \rightarrow v_{c_1}$ path requires three stages but has a cumulative weight of 6 due to the high likelihood of exploiting $v_{c_1}$ from $v_{l_4}$ (0.5) and $v_{l_4}$ from $v_{l_1}$ (0.33).

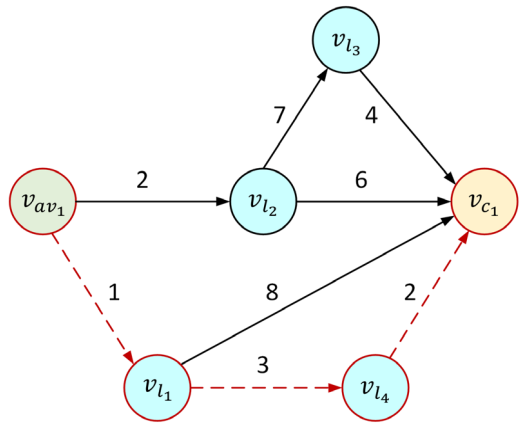

**Fig. 6.** Most feasible/attractive path for threat actors

The shortest path from an attack vector leading to the adversarial consequence has the highest risk score from the manufacturer's point of view. While finding such a path is critical for risk assessment and mitigation, a large number of interconnected components in a manufacturing system yielding many vertices and edges makes manually finding such paths impractical. There are several graph-theoretic algorithms to determine the shortest path between specific vertices. Two commonly used algorithms are the Bellman-Ford algorithm [68] and Dijkstra's algorithm [69]. While the Bellman-Ford algorithm can work on graphs with negative edge weights, it has less scalability and more overhead than Dijkstra's algorithm. Its time complexity is $O(|V||E|)$ that is greater than the time complexity for Dijkstra's algorithm, i.e., $O(|E|\log |V|)$. Considering potential large-scale graphs for smart manufacturing systems and the absence of negative edge weights (as probabilities cannot be negative) in attack graphs, this work adopts Dijkstra's algorithm to identify such shortest paths between two given vertices in the attack graph. Although Dijkstra's algorithm is widely used in network communication systems, this work is the first to use it in manufacturing cybersecurity risk assessment. The pseudo-code of the used algorithm and a brief explanation is given below.

| Dijkstra's algorithm for finding the shortest attack path | |
|---|---|
| **Input:** | Graph: $G$, Starting vertex: $A$ |
| **Output:** | Shortest path and distance |
| 1 | **for** all $u \in V$: |
| 2 | $\quad distance[u] \leftarrow \infty, prev[u] \leftarrow nil$ |
| 3 | $distance[A] \leftarrow 0$ |
| 4 | $H \leftarrow MakeQueue(V)$ {distance-values as keys} |
| 5 | **while** $H$ is not empty: |
| 6 | $\quad u \leftarrow ExtractMin(H)$ |
| 7 | $\quad$ **for** all $(u,v) \in E$: |
| 8 | $\quad\quad$ **if** $distance[v] > distance[u] + w(u,v)$: |
| 9 | $\quad\quad\quad dist[v] \leftarrow dist[u] + w(u,v)$ |
| 10 | $\quad\quad\quad prev[v] \leftarrow u$ |
| 11 | $\quad\quad\quad ChangePriority(H, v, distance[v])$ |

Lines 1 through 3. All vertices in discovered attack paths are considered unvisited, and an enormous distance value is initially assigned to every unvisited vertex. The distance to the starting vertex is considered zero.

Lines 4. An ordered queue is formed with all the vertices based on their distance values. Note that the starting vertex will be the first in the queue during the program initiation as it has the minimum assigned distance value.

Lines 5 through 11. The vertex with the minimum assigned distance value is selected from the queue and considered the current vertex. The distances between the current vertex and the vertices connected to it are calculated. The newly calculated distance is compared to the previously assigned distance value; the smaller of the two values is allocated, and the distance values are updated. After marking all its unvisited neighbors, the current vertex is removed from the queue of unvisited vertices. Then, a new current vertex is identified from the queue, and the process is repeated. This algorithm can also be terminated if the desired target vertex is found. Dijkstra's algorithm can find attack paths with the shortest distance, i.e., the path with the highest probability of compromising the attack locations, from a specific attack vector leading to a potential consequence.

### 3.3.5 Risk estimation

We can identify the shortest attack path using Dijkstra's algorithm. Assume the shortest attack path starts with the attack vector $v_{av_i}$, which initiates the threat event. $v_\lambda$ is the subset of the remaining vertices ($v_\lambda \subset v_l$) representing diverse manufacturing entities that threat actors can access and compromise in the attack path before reaching the attack consequence $v_{c_k}$. $\{v_\lambda^\beta\}_{\beta \subset \{1,...,n\}}$ represents the order of those vertices present in the attack path under consideration involving total $n$ assets. Note that, an attack path passes through these $n$ assets out of the total $N_L$ attack locations. $n$ and $\{v_\lambda^\beta\}_{\beta \subset \{1,...,n\}}$ will be outputs of the used algorithm once manufacturers specify $v_{av_i}$ and $v_{c_k}$, based on the threat attributes applicable to their manufacturing ecosystem. Each edge in the attack path represents different actions of an adversary or cascading effect of the attack vector leading to the final attack impact on the manufacturing system. For example, these sequential actions can represent a threat actor gaining administrator access to the cloud storage containing digital files, tampering with the G code or the tool path files, and altering the Geometric Dimensioning and Tolerancing (GD&T) information to compromise the inspection system, all of which finally result in degraded product quality. Therefore, risks of potential attack paths can be calculated from the probabilities of compromising individual attack locations and expected losses associated with the consequence in the attack graph. So, the risk of an attack path can be calculated as follows:

$$R = P_{AV_i L_j} \left( \prod_{m=1}^{n-1} T_{v_\lambda^m, v_\lambda^{m+1}} \right) T_{v_\lambda^n, v_{c_k}} C_k$$

$P_{AV_i,L_j}$ denotes the probability of compromising the attack location $v_{l_j}$ using attack vector $v_{av_i}$. $T_{v_\lambda^m,v_\lambda^{m+1}}$ represents attack propagation from the vertex $v_\lambda^m$ to vertex $v_\lambda^{m+1}$ in the attack path, $n$ is the total number of attack locations compromised during the threat event. $T_{v_\lambda^n,v_{c_k}}$ is the transition probability from the last vertex of the $\{v_\lambda^\beta\}_{\beta \subset \{1,\dots,n\}}$ subsequence to attack consequence $v_{c_k}$. $C_k$ is the associated consequence representing the lost value in monetary units. The NIST CVSS scoring system recommends quantifying attack consequences in terms of losses due to (1) theft of confidential information, (2) compromised system integrity, and (3) system unavailability. In the manufacturing domain, more specific consequences on business can be evaluated, such as (1) lost sales, (2) increased production waste, (3) recovery costs from sabotage or system damage, (4) cost of operational downtime, (5) repair costs from machine breakdown, (6) safety hazard to personnel, and (7) product-related damages (including reputation damage from degraded product quality and reliability).

# 4 ILLUSTRATIVE EXAMPLE

This section provides an example to illustrate and demonstrate the risk assessment process using the proposed graph-theoretic framework. Note that this illustrative example aims to show how the attack graph formalism and the taxonomy-driven threat characterization can help risk modeling and assessment in smart manufacturing systems. Therefore, the exploitation and attack propagation probabilities are assumed to be known; estimating the probability is out of the scope of this work.

**System description.** This illustrative example shows the cyber-physical manufacturing system of a medium-sized manufacturing organization working as a Manufacturing-as-a-Service (MaaS) provider. MaaS represents a more flexible and digitized approach to manufacturing fostered by Industry 4.0 and the digital transformation of industries, providing manufacturing capabilities and resources on demand. MaaS can also involve a network of suppliers and partners collaborating to fulfill customer orders efficiently. In this example, customers can upload and submit the design file (e.g., CAD or .STL files) to the manufacturer with specific GD&T requirements through a web portal, and those digital files are then stored in the cloud. The manufacturer produces some components in-house and outsources the rest; therefore, product and process data are shared across the manufacturing supply chain with suppliers and vendors. The manufacturer uses a hybrid CNC machine with subtractive and additive manufacturing capabilities on the production floor, which is connected to the internet. The hybrid machine automatically updates the firmware whenever a new version becomes available. A sensor suite monitors production processes in real-time and sends the collected data to the inspection system for post-production quality check. During the pre-defined quality inspection, several key product quality characteristics are monitored, such as dimensions, locations of geometric features, and surface finish. The observed results are compared with the GD&T information stored in the cloud. A worker monitors the system status through a human-machine interface and adjusts the machine set-up if necessary. Finished products are shipped to the customer after the post-production inspection.

**Cybersecurity risk model, assumptions, and attack graph formulation.** The manufacturer is interested in assessing the system's cybersecurity risk posture and identifying the critical assets requiring prioritized security control to defend against cyberattacks. First, the manufacturer needs to identify and characterize potential cybersecurity threats comprising varying attack attributes and potential attack consequences. In doing so, the taxonomical classification of cybersecurity threat attributes and attack consequences presented in Section 3.1 are utilized to populate the nodes of the cyber-physical attack graph. The manufacturer identifies that the supply chain, network communication system, cloud storage, firmware, inspection system, hybrid CNC machine, sensors and actuators, and personnel (the machine operator) are the potential attack locations that fit their manufacturing ecosystem. Next, depending on the specific product line and leveraging expert knowledge, the manufacturer narrows down the attack vectors most relevant to consider for their tailored production ecosystem and the consequences they are concerned about. The organization is currently producing different types of engine brackets, which are used in mission-critical systems. Considering the critical application area, the manufacturer is concerned about outgoing products' geometric integrity and quality. Primarily, two attack vectors are under consideration: a) hardware tampering and b) Man-in-the-Middle attack. The manufacturer assumes threat actors can use network scanning, social engineering techniques, open-source intelligence gathering, research and reconnaissance, insider knowledge, and shared information among supply chain partners to gain information about the target manufacturing system and potential attack locations.

Next, the manufacturer develops an attack graph by adopting the cyberattack modeling and representation scheme described in Section 3.2. The corresponding attack graph consisting of different viable sequence(s) of adversarial actions comprising different attack paths through the system is presented in Fig. 7. The manufacturer uses expert knowledge to identify possible entry points based on available manufacturing assets, existing system vulnerabilities, and the considered attack vectors. For example, the Man-in-the-Middle attack targets digital communication systems, while the physical tampering attack can only target physical assets. Such domain knowledge can facilitate defining the edges from attack vectors to the targeted components in the system. The edges to connect different attack locations together and to potential consequences are defined based on the material and information flow in the system. Additionally, prior attack cases, community reports for identifying and understanding attacks (e.g., CAPEC [70]), academic research, industry reports, and penetration testing can aid in defining those connections.

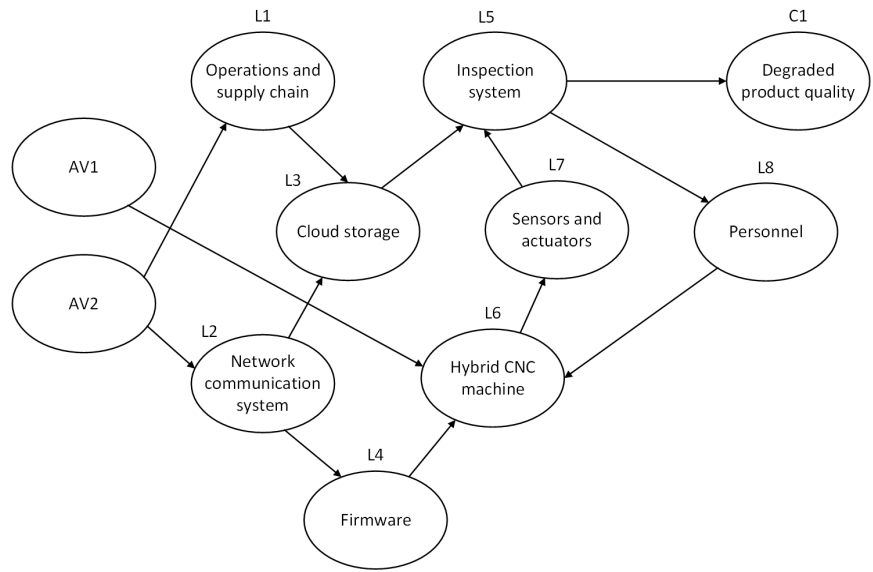

**Fig. 7.** Attack graph for the connected smart manufacturing system considered in the illustrative example

Threat actors can target the hybrid CNC machine to tamper with the hardware ($AV1$) by gaining physical access to the production facility. They can manipulate process parameters (e.g., the cutting speed and coolant usage in machining, and material feed rate and printing speed in the additive manufacturing process) to affect the geometric and structural integrity of the manufactured product. If the sensor suit is compromised or the data collection process is tampered with, the inspection system cannot detect the induced malicious changes [72]. Adversaries can also launch a Man-in-the-Middle attack ($AV2$) targeting the network communication system of the manufacturer and/or any other entities in the manufacturing supply chain network. This attack vector intercepts network traffic and eavesdrops on or manipulates the transmitted data. For example, adversaries can eavesdrop on communications that include login attempts, form submissions, or other data transmission and capture plaintext usernames and passwords. Such stolen credentials can grant them access to the cloud storage where product design files and GD&T specifications are stored. With access to these digital files, threat actors can modify the tool path files or .STL files and introduce defects (such as internal voids), which can significantly degrade the strength of the final product and cause premature failure [8]. Threat actors can also use the stolen credentials to acquire the necessary knowledge about the post-production inspection strategy of a manufacturer and utilize the information about the inspection scheme to design product-oriented attacks that can evade detection during inspection [11]. For example, they can tamper with the part's non-key quality characteristics (KQCs) that are not inspected, alter the GD&T information used for post-production and the definition of a non-conforming product inspection so that the automated quality inspection cannot detect malicious changes, and design physics-informed attacks that do not change commonly used process descriptors (e.g., mean value of a process variable) [9–11]. A compromised inspection system with altered GD&T specifications can also show false system status in the Human-Machine Interface (HMI) system, instigating the personnel to take wrong corrective actions. Additionally, after gaining access to the network system, adversaries can also spoof the address of the public firmware repository that the hybrid CNC machine is designed to reach out checking for updates in the machine firmware and host malicious firmware at the spoofed address [71]. The machine will be compromised once the firmware is updated with the tampered one.

**Attack propagation and attack path discovery.** Following the methodology presented in Section 3.3.1, the attack propagation from potential attack vectors was identified using the DFS algorithm, as shown via the dashed lines in Fig. 8. The starting vertex of the attack path is marked with the star symbol, and rectangle vertices denote the accessible locations leveraging the interdependence between manufacturing assets, and solid lines denote the unexplored edges in the attack graph. There is only one attack path from attack vector $AV1$ to consequence $C1$. However, multiple attack paths can lead to $C1$ from $AV2$. Attack propagation discovers viable attack targets and attack paths representing affected system components. The first takeaway from this analysis is that $AV2$ can potentially compromise more assets than $AV1$, and it has multiple ways to achieve the desired consequence.

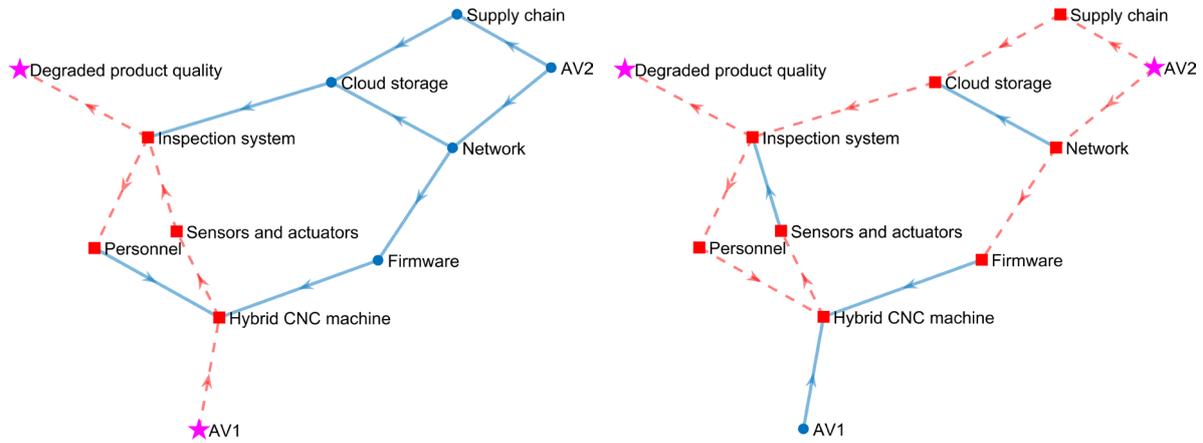

**Fig. 8.** Attack propagation from attack vectors $AV1$ and $AV2$

**Attack propagation likelihood calculation**. The likelihood of compromising a specific asset will differ between attack vectors, depending on attack complexity, privilege required, user interaction, and implemented remediation level. The likelihood of compromising specific manufacturing assets and attack propagation can be computed using the formalism presented in Sections 3.3.2 and 3.3.3. The AV, AC, PR, UI, and RL metrics can be used to estimate the likelihood of compromising specific attack locations by designated attack vectors. For example, physical tampering with the machine ($AV_1L_6$) will require direct access to the targeted machine, which restricts the number of potential threat actors relative to remote attacks that exploit network vulnerabilities. Therefore, the AV score for physical manipulation would be lower than that for network-level attack vectors. For getting physical access to a well-protected device, for instance, it may be necessary to overcome multiple layers of security measures. Hence, physical tampering attacks will have a higher AC score, considering their high complexity. Physical tampering with the machine also requires physical proximity and potentially elevated privileges, such as administrative control. As a result, it will receive higher PR scores than those that require fewer privileges. However, the UI score for physical tampering attacks will be comparatively lower as they rely less on user interaction. Finally, the RL metric evaluates the required remediation effort to reduce the risk of the physical tampering attack vector. Implementing physical security measures, such as surveillance cameras, access controls, and tamper-evident seals, can reduce the likelihood of physical interference with the machine. Therefore, a greater RL score indicates a lower likelihood of successful physical tampering attacks. Thus, the organization can determine the probability of compromising different attack locations by specific attack vectors. Similarly, the attack propagation likelihood can be specified based on the configuration and topology of manufacturing assets, data flow map in the organization, historical data, and expert knowledge. For the demonstration purpose, assume that the probabilities of compromising different specific attack locations and the attack propagation through the considered manufacturing system are given in Table 1. With the attack graph defined, we are now interested in analyzing the attack propagation, finding the shortest attack path, and computing the risk.

**Table 1:** Sample probabilities for compromising an attack location and attack propagation and the respective weights for the graph edges

| Edges | $AV_1L_6$ | $AV_2L_1$ | $AV_2L_2$ | $L_1L_3$ | $L_2L_3$ | $L_2L_4$ | $L_3L_5$ | $L_4L_6$ | $L_5L_8$ | $L_5C_1$ | $L_6L_7$ | $L_7L_5$ | $L_8L_6$ |
|---|---|---|---|---|---|---|---|---|---|---|---|---|---|
| Probability | 0.2 | 0.35 | 0.6 | 0.15 | 0.3 | 0.3 | 0.25 | 0.9 | 0.05 | 0.8 | 0.05 | 0.6 | 0.3 |
| Weight | 5.00 | 2.86 | 1.67 | 6.67 | 3.33 | 3.33 | 4.00 | 1.11 | 20.00 | 1.25 | 20.00 | 1.67 | 3.33 |

**Shortest attack path and risk estimation.** The shortest attack path and the associated risk are determined following the method presented in Sections 3.3.4 and 3.3.5. The shortest path represents the preferred attack sequence from the perspective of threat actors involving the minimal detection probability by manufacturers. From the manufacturer's point of view, this attack sequence translates to the most vulnerable connections and critical manufacturing assets requiring prioritized security control and detective and protective defense measures. The identified shortest path from $AV2$ to $C1$ will constitute the highest risk for the manufacturer, highlighting the most critical manufacturing assets requiring protection. Dijkstra's algorithm was used to find the shortest attack path, and the result is presented in Fig. 9. The algorithm provides the shortest path, a list of vertices within the attack path, and the number of steps in the attack sequence (hop length). In the current system setting, $AV2 \rightarrow L2 \rightarrow L3 \rightarrow L5 \rightarrow C1$ is the most attractive attack path for adversaries. The cumulative risks for all possible attack paths were calculated, which are listed in Table 2. Here, $C$ can be defined in monetary terms by the organization. The cumulative risk calculation also verifies the result obtained from the algorithm.

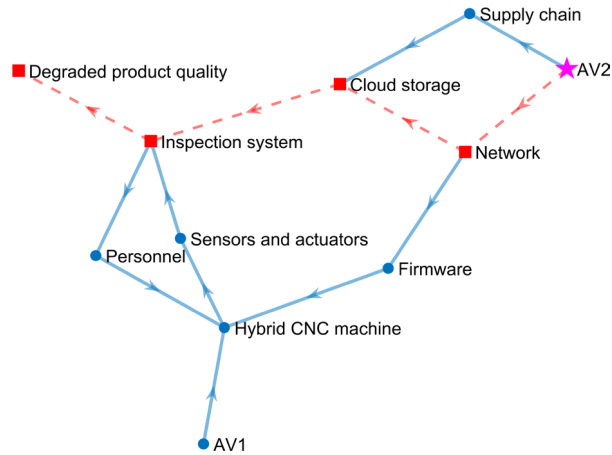

**Fig. 9.** Shortest attack path (attack path with the highest cumulative risk) between $AV2$ and $C1$

**Table 2.** Risk calculation for different attack paths

| Path no. | Attack path from $AV2$ | Hop length | Cumulative Risk |
|----------|------------------------|------------|-----------------|
| 1a | $AV2 \rightarrow L1 \rightarrow L3 \rightarrow L5 \rightarrow C1$ | 4 | $0.0105 \times C$ |
| 2a | $AV2 \rightarrow L2 \rightarrow L3 \rightarrow L5 \rightarrow C1$ | 4 | $\mathbf{0.036 \times C}$ |
| 3a | $AV2 \rightarrow L2 \rightarrow L4 \rightarrow L6 \rightarrow L7 \rightarrow L5 \rightarrow C1$ | 6 | $0.0039 \times C$ |

The likelihood of compromising attack locations and attack propagation can change based on new vulnerabilities and/or deployed defense measures. The updated likelihood will also be reflected in the edge weights in the attack graph and can modify the shortest attack path. For example, manufacturers may deploy an intrusion prevention system, authentication measures, and proper access control, decreasing the probability of unauthorized access to the cloud storage through the network communication system and the supply chain network. As a result, attack paths passing through the cloud storage will become more challenging for threat actors due to the high detection probability. Table 3 represents the tentatively updated probabilities and edge weights for the attack graph after considering new vulnerabilities and defenses.

**Table 3.** Updated probabilities for compromising an attack location and attack propagation and the respective weights for the graph edges

| Edges | $AV_1L_6$ | $AV_2L_1$ | $AV_2L_2$ | $L_1L_3$ | $L_2L_3$ | $L_2L_4$ | $L_3L_5$ | $L_4L_6$ | $L_5L_8$ | $L_5C_1$ | $L_6L_7$ | $L_7L_5$ | $L_8L_6$ |
|-------|-----------|-----------|-----------|----------|----------|----------|----------|----------|----------|----------|----------|----------|----------|
| Probability | 0.2 | 0.35 | 0.6 | .05 | .05 | 0.3 | .05 | 0.1 | 0.05 | 0.8 | 0.6 | 0.9 | 0.3 |
| Weight | 5.00 | 2.86 | 1.67 | 20.00 | 20.00 | 3.33 | 20.00 | 10.00 | 20.00 | 1.25 | 1.67 | 1.11 | 3.33 |

For the updated probabilities, the algorithm identified $AV2 \rightarrow L2 \rightarrow L4 \rightarrow L6 \rightarrow L7 \rightarrow L5 \rightarrow C1$ as the new shortest attack path, which is illustrated in Fig. 10. The cumulative risks for all possible attack paths are listed in Table 4, verifying the result obtained from the algorithm. The most feasible path for threat actors is now bypassing cloud storage. Similarly, the shortest attack path can be updated whenever there is a change in exploitation probabilities. The value of the obtained result is twofold: 1) the identified shortest path defines the status quo of the most critical connections and assets in the manufacturing value chain, and 2) the cumulative risk metric refers to the potential risk posture of the system. Implementing additional defense measures reduced the maximum risk score from $0.036 \times C$ to $0.0078 \times C$. Manufacturers may accept the risk up to a specific threshold value depending on the type of organization. Nevertheless, improving the existing defense scheme or developing new ones at the critical connections and assets in the shortest attack path will further reduce the risk of cyberattacks on the system.

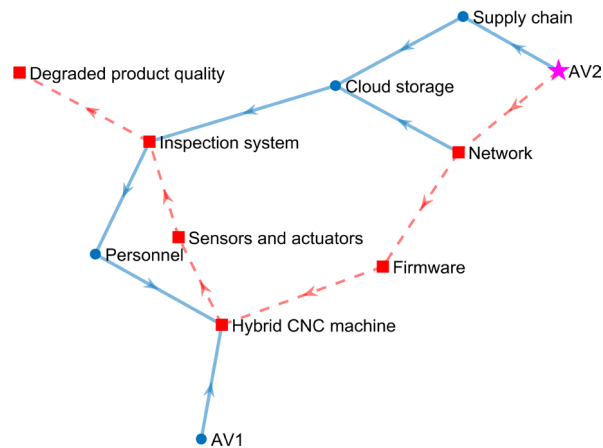

**Fig. 10.** The updated shortest attack path (path with the highest cumulative risk) between $AV2$ and $C1$

**Table 4.** Updated risk calculation for different attack paths

| Path no. | Attack path from $AV2$ | Hop length | Cumulative Risk |
|---|---|---|---|
| 1b | $AV2 \rightarrow L1 \rightarrow L3 \rightarrow L5 \rightarrow C1$ | 4 | $0.0007 \times C$ |
| 2b | $AV2 \rightarrow L2 \rightarrow L3 \rightarrow L5 \rightarrow C1$ | 4 | $0.0012 \times C$ |
| 3b | $AV2 \rightarrow L2 \rightarrow L4 \rightarrow L6 \rightarrow L7 \rightarrow L5 \rightarrow C1$ | 6 | $\mathbf{0.0078 \times C}$ |

**Discussion and takeaways**. The illustrative example demonstrates how the proposed cybersecurity risk assessment framework can be tailored to the specific characteristics of a manufacturing ecosystem. The general manufacturing value chain presented in the example also applies to numerous small and medium-sized manufacturers emerging as Manufacturing-as-a-service providers. Additionally, manufacturers with different sets of manufacturing assets that may have specific vulnerabilities, attack vectors, and potential consequences can still apply the proposed methodology similarly. Although some modern manufacturing systems may scale up significantly with numerous manufacturing assets and processes, manufacturers can conduct a more focused risk assessment by only considering the relevant attack vectors and attack consequences of concern. They will primarily analyze subgraphs for specific attack vectors and consequences, even if the entire attack graph is complex. However, manufacturers using mobile assets to handle different products on the shop floor may trigger dynamic changes in the attack graph. In such cases, different instances of the dynamic graph structure can be enumerated based on the product line and process planning, which will have a finite number of attack graph scenarios. Each scenario can have individual risk profiles, and the methodology proposed in this work will still apply. It is also worth mentioning that the layout and configuration of manufacturing systems do not change significantly or very often in practice, leading to static attack graphs. Securing the system across different attack graph combinations that change in real time will require expanding the current work, which is beyond the scope of this paper.

Combining the attack graph formalism with taxonomy-driven threat characterization, the proposed framework ensures that the generated attack graphs are consistent and comprehensive. When the exploitation and attack propagation probabilities are defined, which were assumed in this example, manufacturers can visualize potential attack paths stemming from specific attack vectors, assess the cascading effects, evaluate the associated risks, and identify the critical attack paths with the highest risk. Changes in the network topology and probability measures affect the critical attack paths, and the proposed approach can capture them. The proposed approach requires periodic updating and reidentifying critical attack paths posing the highest risk when new assets are included in the discrete manufacturing system, new vulnerabilities are identified, and/or novel defenses are deployed.

# 5 CONCLUSION

To secure critical manufacturing infrastructure against high stakes cyberattacks, manufacturing stakeholders need a proactive risk management approach. This is key to identifying relevant security threats to the organization, cyber-physical infrastructure vulnerabilities, the likelihood of threat events, potential impacts, and mitigation strategies. Graph-based risk assessment approaches have been effective and widely used in various cyber-physical systems to manage security risks. However, existing approaches focus on limited attack methods and consequences, which cannot represent the convoluted risk landscape in discrete manufacturing systems. Systematic generation of comprehensive attack graphs for the discrete manufacturing system is another significant challenge in utilizing the attack graph formalism. Most importantly, the literature lacks a tailored graph-based risk modeling and assessment framework specifically designed for the unique challenges and vulnerabilities present in discrete manufacturing systems. In response, this paper introduces the first graph-based formal model and framework to represent the cybersecurity threat landscape in discrete manufacturing systems, analyze attack propagation, visualize potential attack paths, compute the security risk for the identified attack paths, evaluate the attack path with the highest risk to manufacturers, and identify vulnerable manufacturing assets requiring prioritized control.

Our proposed framework uses attack graph formalism to incorporate different threat attributes into the risk model for facilitating the concurrent modeling and analysis of various cybersecurity threats comprising varying attack attributes. In doing so, we adopt manufacturing-specific taxonomical classifications of cyber-physical attack vectors, locations, vulnerabilities, and consequences for systematic and comprehensive characterization of the tactics, techniques, and procedures that threat actors use. Those taxonomical classifications are then used to generate comprehensive and generalizable cyber-physical attack graphs. We use attack graphs to model the cascading attack impact through different cyber and physical entities in manufacturing systems, leading to specific consequences. The constructed attack graphs are then analyzed using the DFS algorithm to identify potential attack paths, showing how threat events originating from different attack vectors can propagate through the manufacturing value chain to realize specific consequences. This enables identifying manufacturing assets that threat actors can access and compromise during potential threat events. We also present a quantitative model to estimate the cybersecurity risk associated with the identified attack paths. Then, we utilize Dijkstra's algorithm to find the attack path posing the highest risk and identify the most critical manufacturing assets. The risk model proposed in this work assumes that threat actors targeting manufacturing systems will have detailed knowledge of the system, possess broad attack capabilities, and prioritize ensuring an attack's success regardless of the associated cost and time required. However, there can be a trade-off between the cost and time requirement for launching specific attacks and their likelihood to succeed from the threat actor's perspective, which may influence their attack path selection. Considering such a trade-off is beyond the scope of the current work and is worth further analysis. As a potential future direction, we consider focusing on deriving the exploitation and attack propagation probabilities based on real case studies and/or simulation results. Risk mitigation strategies and recovery plans during threat events will also be investigated using graph-theoretic tools, such as isolating a compromised manufacturing asset from the network using the minimum cut theorem to minimize attack propagation instead of shutting down the entire system. Our proposed framework offers a critical step toward aiding researchers and practitioners in identifying the most critical connections and assets in a discrete manufacturing system. It also helps in showing how those assets can be exploited by different attack vectors, informing decision-makers to prioritize security investments for developing and deploying appropriate cybersecurity defense measures.

## FUNDING

This research was partially funded by Arizona's Technology and Research Initiative Fund (TRIF) under the National Security Systems Initiative.

## REFERENCES

[1]   Lu, Y., Morris, K. C., and Frechette, S., 2016, *Current Standards Landscape for Smart Manufacturing Systems*.
[2]   Lu, Y., Xu, X., and Wang, L., 2020, "Smart Manufacturing Process and System Automation–a Critical Review of the Standards and Envisioned Scenarios," J. Manuf. Syst., **56**, pp. 312–325.
[3]   Tweneboah-Koduah, S., Skouby, K. E., and Tadayoni, R., 2017, "Cyber Security Threats to IoT Applications and Service Domains," Wirel. Pers. Commun., **95**(1), pp. 169–185.

[4]     Roman, R., Najera, P., and Lopez, J., 2011, "Securing the Internet of Things," Computer (Long. Beach. Calif)., (9), pp. 51–58.

[5]     Da Xu, L., He, W., and Li, S., 2014, "Internet of Things in Industries: A Survey," IEEE Trans. Ind. informatics, **10**(4), pp. 2233–2243.

[6]     Rahman, M. H., Wuest, T., and Shafae, M., 2023, "Manufacturing Cybersecurity Threat Attributes and Countermeasures: Review, Meta-Taxonomy, and Use Cases of Cyberattack Taxonomies," J. Manuf. Syst., **68**, pp. 196–208.

[7]     2020, "IBM Security X-Force Threat Intelligence Index" [Online]. Available: https://securityintelligence.com/posts/threat-actors-targeted-industries-2020-finance-manufacturing-energy/. [Accessed: 08-Mar-2022].

[8]     Sturm, L. D., Williams, C. B., Camelio, J. A., White, J., and Parker, R., 2017, "Cyber-Physical Vulnerabilities in Additive Manufacturing Systems: A Case Study Attack on the. STL File with Human Subjects," J. Manuf. Syst., **44**, pp. 154–164.

[9]     Elhabashy, A. E., Wells, L. J., Camelio, J. A., and Woodall, W. H., 2019, "A Cyber-Physical Attack Taxonomy for Production Systems: A Quality Control Perspective," J. Intell. Manuf., **30**(6), pp. 2489–2504.

[10]    Rahman, M. H., and Shafae, M., 2022, "Physics-Based Detection of Cyber-Attacks in Manufacturing Systems: A Machining Case Study," J. Manuf. Syst., **64**, pp. 676–683.

[11]    Shafae, M. S., Wells, L. J., and Purdy, G. T., 2019, "Defending against Product-Oriented Cyber-Physical Attacks on Machining Systems," Int. J. Adv. Manuf. Technol., pp. 1–21.

[12]    Mahesh, P., Tiwari, A., Jin, C., Kumar, P. R., Reddy, A. L. N., Bukkapatanam, S. T. S., Gupta, N., and Karri, R., 2021, "A Survey of Cybersecurity of Digital Manufacturing," Proc. IEEE, **109**(4), pp. 495–516.

[13]    Tuptuk, N., and Hailes, S., 2018, "Security of Smart Manufacturing Systems," J. Manuf. Syst., **47**, pp. 93–106.

[14]    Wu, D., Ren, A., Zhang, W., Fan, F., Liu, P., Fu, X., and Terpenny, J., 2018, "Cybersecurity for Digital Manufacturing," J. Manuf. Syst., **48**, pp. 3–12.

[15]    Turner, H., White, J., Camelio, J. A., Williams, C., Amos, B., and Parker, R., 2015, "Bad Parts: Are Our Manufacturing Systems at Risk of Silent Cyberattacks?," IEEE Secur. Priv., **13**(3), pp. 40–47.

[16]    Graves, L. M. G., King, W., Carrion, P., Shao, S., Shamsaei, N., and Yampolskiy, M., 2021, "Sabotaging Metal Additive Manufacturing: Powder Delivery System Manipulation and Material-Dependent Effects," Addit. Manuf., p. 102029.

[17]    Chhetri, S. R., Canedo, A., and Faruque, M. A. Al, 2017, "Confidentiality Breach through Acoustic Side-Channel in Cyber-Physical Additive Manufacturing Systems," ACM Trans. Cyber-Physical Syst., **2**(1), pp. 1–25.

[18]    Wu, M., Song, Z., and Moon, Y. B., 2019, "Detecting Cyber-Physical Attacks in CyberManufacturing Systems with Machine Learning Methods," J. Intell. Manuf., **30**(3), pp. 1111–1123.

[19]    Belikovetsky, S., Solewicz, Y. A., Yampolskiy, M., Toh, J., and Elovici, Y., 2019, "Digital Audio Signature for 3D Printing Integrity," IEEE Trans. Inf. Forensics Secur., **14**(5), pp. 1127–1141.

[20]    Komolafe, T., Tian, W., Purdy, G. T., Albakri, M., Tarazaga, P., and Camelio, J., 2019, "Repeatable Part Authentication Using Impedance Based Analysis for Side-Channel Monitoring," J. Manuf. Syst., **51**, pp. 42–51.

[21]    2011, *NIST SP 800-39. Managing Information Security Risk: Organization, Mission, and Information System View*, National Institute of Standards & Technology.

[22]    Blank, R. M., and Gallagher, P. D., 2012, *Guide for Conducting Risk Assessments*.

[23]    Stouffer, K., Zimmerman, T., Tang, C., Lubell, J., Cichonski, J., and Mccarthy, J., 2020, *NISTIR 8183 Revision 1, Cybersecurity Framework: Manufacturing Profile*.

[24]    2022, "CyManII Roadmap" [Online]. Available: https://www.energy.gov/eere/articles/does-cybersecurity-manufacturing-innovation-institute-releases-first-public-roadmap. [Accessed: 20-Dec-2022].

[25]    Liu, H.-C., You, J.-X., Chen, S., and Chen, Y.-Z., 2016, "An Integrated Failure Mode and Effect Analysis Approach for Accurate Risk Assessment under Uncertainty," IIE Trans., **48**(11), pp. 1027–1042.

[26]    Sherwin, M. D., Medal, H. R., MacKenzie, C. A., and Brown, K. J., 2020, "Identifying and Mitigating Supply Chain Risks Using Fault Tree Optimization," IISE Trans., **52**(2), pp. 236–254.

[27]    Poolsappasit, N., Dewri, R., and Ray, I., 2011, "Dynamic Security Risk Management Using Bayesian Attack Graphs," IEEE Trans. Dependable Secur. Comput., **9**(1), pp. 61–74.

[28]    Sen, A., and Madria, S., 2016, "Risk Assessment in a Sensor Cloud Framework Using Attack Graphs," IEEE Trans. Serv. Comput., **10**(6), pp. 942–955.

[29]    Ge, M., Hong, J. B., Guttmann, W., and Kim, D. S., 2017, "A Framework for Automating Security

Analysis of the Internet of Things," J. Netw. Comput. Appl., **83**, pp. 12–27.

[30] Huang, K., Zhou, C., Tian, Y.-C. C., Yang, S., and Qin, Y., 2018, "Assessing the Physical Impact of Cyberattacks on Industrial Cyber-Physical Systems," IEEE Trans. Ind. Electron., **65**(10), pp. 8153–8162.

[31] Lyu, X., Ding, Y., and Yang, S.-H. H., 2020, "Bayesian Network Based C2P Risk Assessment for Cyber-Physical Systems," IEEE Access, **8**, pp. 88506–88517.

[32] Chae, Y. H., Lee, C., Choi, M. K., and Seong, P. H., 2022, "Evaluating Attractiveness of Cyberattack Path Using Resistance Concept and Page-Rank Algorithm," Ann. Nucl. Energy, **166**, p. 108748.

[33] Jha, S., Sheyner, O., and Wing, J., 2002, "Two Formal Analyses of Attack Graphs," *Proceedings 15th IEEE Computer Security Foundations Workshop. CSFW-15*, IEEE, pp. 49–63.

[34] Ou, X., Govindavajhala, S., and Appel, A. W., 2005, "MulVAL: A Logic-Based Network Security Analyzer.," *USENIX Security Symposium*, Baltimore, MD, pp. 113–128.

[35] Ingols, K., Chu, M., Lippmann, R., Webster, S., and Boyer, S., 2009, "Modeling Modern Network Attacks and Countermeasures Using Attack Graphs," *2009 Annual Computer Security Applications Conference*, IEEE, pp. 117–126.

[36] Jia, F., Hong, J. B., and Kim, D. S., 2015, "Towards Automated Generation and Visualization of Hierarchical Attack Representation Models," *2015 IEEE International Conference on Computer and Information Technology; Ubiquitous Computing and Communications; Dependable, Autonomic and Secure Computing; Pervasive Intelligence and Computing*, IEEE, pp. 1689–1696.

[37] Cai, C., Zhang, Y., Wang, Z., and Xue, C., 2019, "A New Model for Securing Networks Based on Attack Graph," 2019 IEEE 4th Int. Conf. Signal Image Process. ICSIP 2019, pp. 318–324.

[38] Wu, W., Kang, R., and Li, Z., 2016, "Risk Assessment Method for Cybersecurity of Cyber-Physical Systems Based on Inter-Dependency of Vulnerabilities," IEEE Int. Conf. Ind. Eng. Eng. Manag., **2016-Janua**, pp. 1618–1622.

[39] George, G., and Thampi, S. M., 2018, "A Graph-Based Security Framework for Securing Industrial IoT Networks from Vulnerability Exploitations," IEEE Access, **6**, pp. 43586–43601.

[40] Al Ghazo, A. T., Ibrahim, M., Ren, H., and Kumar, R., 2019, "A2G2V: Automatic Attack Graph Generation and Visualization and Its Applications to Computer and SCADA Networks," IEEE Trans. Syst. Man, Cybern. Syst., **50**(10), pp. 3488–3498.

[41] Ani, U. D., He, H., and Tiwari, A., 2020, "Vulnerability-Based Impact Criticality Estimation for Industrial Control Systems," Int. Conf. Cyber Secur. Prot. Digit. Serv. Cyber Secur. 2020.

[42] Stergiopoulos, G., Dedousis, P., and Gritzalis, D., 2022, "Automatic Analysis of Attack Graphs for Risk Mitigation and Prioritization on Large-Scale and Complex Networks in Industry 4.0," Int. J. Inf. Secur., pp. 1–23.

[43] Elhabashy, A. E., Wells, L. J., and Camelio, J. A., 2020, "Cyber-Physical Attack Vulnerabilities in Manufacturing Quality Control Tools," Qual. Eng., **32**(4), pp. 676–692.

[44] DeSmit, Z., Elhabashy, A. E., Wells, L. J., and Camelio, J. A., 2017, "An Approach to Cyber-Physical Vulnerability Assessment for Intelligent Manufacturing Systems," J. Manuf. Syst., **43**, pp. 339–351.

[45] "What Is a Threat Actor? | IBM" [Online]. Available: https://www.ibm.com/topics/threat-actor. [Accessed: 15-Aug-2023].

[46] Sailio, M., Latvala, O.-M., and Szanto, A., 2020, "Cyber Threat Actors for the Factory of the Future," Appl. Sci., **10**(12), p. 4334.

[47] 2023, "2022 ICS/OT Cybersecurity Year in Review | Dragos" [Online]. Available: https://www.dragos.com/blog/industry-news/2022-dragos-year-in-review-now-available/. [Accessed: 23-Feb-2023].

[48] 2020, "Manufacturing Threat Perspective | Dragos" [Online]. Available: https://www.dragos.com/resource/manufacturing-threat-perspective/. [Accessed: 17-Jan-2023].

[49] Ahmed, M., and Pathan, A.-S. K., 2020, "False Data Injection Attack (FDIA): An Overview and New Metrics for Fair Evaluation of Its Countermeasure," Complex Adapt. Syst. Model., **8**(1), pp. 1–14.

[50] Bhushan, B., Sahoo, G., and Rai, A. K., 2017, "Man-in-the-Middle Attack in Wireless and Computer Networking—A Review," *2017 3rd International Conference on Advances in Computing, Communication & Automation (ICACCA)(Fall)*, IEEE, pp. 1–6.

[51] FireEye, "What Is a Zero-Day Exploit? | FireEye" [Online]. Available: https://us.norton.com/blog/emerging-threats/how-do-zero-day-vulnerabilities-work. [Accessed: 05-Dec-2022].

[52] El Abbadi, R., and Jamouli, H., 2021, "Takagi–Sugeno Fuzzy Control for a Nonlinear Networked System Exposed to a Replay Attack," Math. Probl. Eng., **2021**.

[53] Industrial Control Systems Cyber Emergency Response Team, 2016, *Recommended Practice: Improving Industrial Control System Cybersecurity with Defense-in-Depth Strategies*.

[54] "China's Huawei and ZTE Pose National Security Threat, Says US Committee | Technology | The Guardian" [Online]. Available: https://www.theguardian.com/technology/2012/oct/08/china-huawei-zte-security-threat. [Accessed: 19-Jan-2023].

[55] Thornburgh, T., 2004, "Social Engineering: The" Dark Art"," *Proceedings of the 1st Annual Conference on Information Security Curriculum Development*, pp. 133–135.

[56] 2020, "Hackers Could Destroy 3D Printers by Setting Them on Fire | TechRadar" [Online]. Available: https://www.techradar.com/news/hackers-could-destroy-3d-printers-by-setting-them-on-fire. [Accessed: 23-Aug-2023].

[57] 2021, "Colonial Pipeline Cyber Attack: Hackers Used Compromised Password - Bloomberg" [Online]. Available: https://www.bloomberg.com/news/articles/2021-06-04/hackers-breached-colonial-pipeline-using-compromised-password. [Accessed: 27-Jan-2023].

[58] Wells, L. J., Camelio, J. A., Williams, C. B., and White, J., 2014, "Cyber-Physical Security Challenges in Manufacturing Systems," Manuf. Lett., **2**(2), pp. 74–77.

[59] Kaspersky, 2022, "The Human Factor in IT Security: How Employees Are Making Businesses Vulnerable from Within." [Online]. Available: https://www.kaspersky.com/blog/the-human-factor-in-it-security/. [Accessed: 03-Feb-2023].

[60] Al Faruque, M. A., Chhetri, S. R., Canedo, A., and Wan, J., 2016, "Acoustic Side-Channel Attacks on Additive Manufacturing Systems," *2016 ACM/IEEE 7th International Conference on Cyber-Physical Systems, ICCPS 2016 - Proceedings*, IEEE, pp. 1–10.

[61] 2020, "Siemens SPPA-T3000 | CISA" [Online]. Available: https://www.us-cert.gov/ics/advisories/icsa-19-351-02. [Accessed: 08-Aug-2023].

[62] 2023, "IBM Security X-Force Threat Intelligence Index" [Online]. Available: https://www.ibm.com/reports/threat-intelligence. [Accessed: 29-Mar-2023].

[63] 2022, "Toyota Cyberattack: Production to Restart in Japan after Attack on Kojima Industries | CNN Business" [Online]. Available: https://www.cnn.com/2022/03/01/business/toyota-japan-cyberattack-production-restarts-intl-hnk/index.html. [Accessed: 19-Jan-2023].

[64] Yampolskiy, M., King, W. E., Gatlin, J., Belikovetsky, S., Brown, A., Skjellum, A., and Elovici, Y., 2018, "Security of Additive Manufacturing: Attack Taxonomy and Survey," Addit. Manuf., **21**(November 2017), pp. 431–457.

[65] Tarjan, R., 1972, "Depth-First Search and Linear Graph Algorithms," SIAM J. Comput., **1**(2), pp. 146–160.

[66] 2019, "Common Vulnerability Scoring System Version 3.1: Specification Document" [Online]. Available: https://www.first.org/cvss/specification-document. [Accessed: 10-Nov-2022].

[67] Kaspersky, 2021, "Zero-Day Exploits & Zero-Day Attacks" [Online]. Available: https://usa.kaspersky.com/resource-center/definitions/zero-day-exploit. [Accessed: 05-Dec-2022].

[68] Goldberg, A., and Radzik, T., 1993, *A Heuristic Improvement of the Bellman-Ford Algorithm*, STANFORD UNIV CA DEPT OF COMPUTER SCIENCE.

[69] Dijkstra, E. W., 1959, "A Note on Two Problems in Connexion with Graphs," Numer. Math., pp. 269–271.

[70] 2023, "Common Attack Pattern Enumeration and Classification (CAPEC)" [Online]. Available: https://capec.mitre.org/. [Accessed: 28-Jan-2023].

[71] McInerney, D., 2020, "With IoT, Common Devices Pose New Threats," Coalfire [Online]. Available: https://www.coalfire.com/the-coalfire-blog/april-2020/with-iot-common-devices-pose-new-threats. [Accessed: 14-Jan-2023].

[72] Vatanparvar, K., Abdullah, M., and Faruque, A., 2019, *Self-Secured Control with Anomaly Detection and Recovery in Automotive Cyber-Physical Systems*.